\documentclass[fleqn,usenatbib]{mnras}
\usepackage{newtxtext,newtxmath}
\usepackage[T1]{fontenc}
\usepackage{ae,aecompl}

\usepackage{multirow}
\usepackage{graphicx}	
\usepackage{amsmath}	
\usepackage{amssymb}	
\usepackage{etoolbox}
\makeatletter
\patchcmd\@combinedblfloats{\box\@outputbox}{\unvbox\@outputbox}{}{\errmessage{\noexpand patch failed}}
\makeatother

\title[I. Sub-kpc MS in nearby spirals]{A panchromatic spatially-resolved analysis of nearby galaxies - I. Sub-kpc scale Main Sequence in grand-design spirals}

\author[Enia A. et al.]{
A. Enia$^{1,2}$\thanks{E-mail: andrea.enia@unipd.it},
G. Rodighiero$^{1,2}$,
L. Morselli$^{1,2}$,
V. Casasola$^{3,4}$,
S. Bianchi$^{4}$,
\newauthor
L. Rodriguez-Mu{\~n}oz$^{1,2}$,
C. Mancini$^{1,2}$,
A. Renzini$^{2}$, 
P. Popesso$^{5}$,
P. Cassata$^{1,2}$,
\newauthor
M. Negrello$^{6}$,
A. Franceschini$^{1}$
\\
\\
$^{1}$ Dipartimento di Fisica e Astronomia, Universit{\`a} di Padova, vicolo dell'Osservatorio 3, I-35122 Padova, Italy\\
$^{2}$ INAF $-$ Osservatorio Astrofisico di Padova, vicolo dell'Osservatorio 5, I-35122 Padova, Italy\\
$^{3}$ INAF $-$ Istituto di Radioastronomia, Via P. Gobetti 101, 40129, Bologna, Italy\\
$^{4}$ INAF $-$ Osservatorio Astrofisico di Arcetri, Largo E. Fermi 5, 50125, Firenze, Italy\\
$^{5}$ Excellence Cluster Universe, Boltzmannstrasse 2, 85748, Garching bei Munchen, Germany\\
$^{6}$ School of Physics and Astronomy, Cardiff University, The Parade, Cardiff CF24 3AA, UK
}

\date{Accepted XXX. Received YYY; in original form ZZZ}

\pubyear{2020}

\begin{document}
\label{firstpage}
\pagerange{\pageref{firstpage}--\pageref{lastpage}}
\maketitle

\begin{abstract}
We analyse the spatially resolved relation between stellar mass (M$_{\star}$) and star formation rate (SFR) in disk galaxies (i.e. the Main Sequence, MS). The studied sample includes eight nearby face-on grand-design spirals, e.g. the descendant of high-redshift, rotationally-supported star-forming galaxies. We exploit photometric information over 23 bands, from the UV to the far-IR, from the publicly available DustPedia database to build spatially resolved maps of stellar mass and star formation rates on sub-galactic scales of 0.5-1.5 kpc, by performing a spectral energy distribution fitting procedure that accounts for both the observed and the obscured star formation processes, over a wide range of internal galaxy environments (bulges, spiral arms, outskirts). With more than 30 thousands physical cells, we have derived a definition of the local spatially resolved MS per unit area for disks, $\log(\Sigma_{SFR})$=0.82log$(\Sigma_{*})$-8.69. This is consistent with the bulk of recent results based on optical IFU, using the H$\alpha$ line emission as a SFR tracer. Our work extends the analysis at lower sensitivities in both M$_{\star}$ and SFR surface densities, up to a factor $\sim$ 10. The self consistency of the MS relation over different spatial scales, from sub-galactic to galactic, as well as with a rescaled correlation obtained for high redshift galaxies, clearly proves its universality. 
\end{abstract}




\section{Introduction}\label{sec:Introduction}

Galaxies appear to build their stellar masses in a steady mode mainly dominated by secular processes and thanks to the accretion of cold gas. This picture finds its confirmation in the existence of a tight relation between the galaxy stellar mass (M$_{\star}$) and its star formation rate (SFR): the Main Sequence (MS) of star-forming galaxies (SFGs), observed up to z$\sim$6 with a fairly constant scatter of $\sim$0.3 dex \citep[e.g.][]{2007ApJ...660L..47N, 2007A&A...468...33E, 2007ApJ...670..156D, 2009ApJ...698L.116P, 2009A&A...504..751S, 2010MNRAS.405.2279O, 2014MNRAS.443...19R, 2014MNRAS.437.3516S, 2014ApJ...791L..25S, 2014ApJS..214...15S, 2014ApJ...795..104W,2015ApJ...804..149S, 2015A&A...575A..74S, 2015ApJ...801L..29R, 2017ApJ...847...76S, 2018A&A...615A.146P, 2019MNRAS.483.3213P}. Galaxies seem to oscillate around the MS relation as a consequence of multiple events of central compaction of gas followed by inside-out gas depletion, thus related with the flows of cold gas in galaxies \citep{2016MNRAS.457.2790T}. 
Several works in the recent years exploited the MS relation as a reference to understand the differences among galaxies characterised by different rates of stellar production (starbursts, SFGs, passive galaxies), with the final aim of understanding the origins of galaxy bimodality, and how the star formation activity is quenched \citep[e.g.][]{2011ApJ...739L..40R, 2015Natur.521..192P,2016MNRAS.462.1749S}.

The existence of a tight relation between stellar mass surface density ($\Sigma_{\star}$) and star formation rate surface density ($\Sigma_{\rm SFR}$) found in HII regions of nearby galaxies, suggested that the global MS relation originates thanks to local processes that set the conversion of gas into stars \citep{2012ApJ...756L..31R, 2013A&A...554A..58S}. Such observations were complemented by the work of \citet{2013ApJ...779..135W}, that discovered a correlation between $\Sigma_{\star}$ and $\Sigma_{\rm SFR}$ on scales of 1 kpc in galaxies at 0.7<z<1.5 thanks to the combination of multiwavelength broad band imaging from the Cosmic Assembly Near-infrared Deep Extragalactic Legacy Survey \citep[CANDELS,][]{2011ApJS..197...35G, 2011ApJS..197...36K} and 3DHST data \citep{2012ApJS..200...13B}. 

Following this, several works exploited the advent of large integral field spectroscopic (IFS) surveys to analyse the existence of a spatially resolved MS relation at low redshift, using the H$\alpha$ flux as SFR tracer. \citet{2016ApJ...821L..26C} used the Calar Alto Legacy Integral Field Area Survey \citep[CALIFA,][]{2012A&A...538A...8S} galaxies and found a spatially resolved MS on 0.5-1.5kpc scales with a slope of 0.72 $\pm$ 0.04 and a scatter of 0.23 dex. \citet{2017ApJ...851L..24H} identified HII spaxels in MaNGA \citep[Mapping Nearby Galaxies at APO,][]{2015ApJ...798....7B} SFGs and found a linear resolved MS on kpc scales with a scatter 0.127. These results suggest that the global relation is set locally, possibly by the surface density of molecular gas \citep{2017ApJ...851...18L}. Exploiting pixel-by-pixel Spectral Energy Distribution (SED) fitting, \cite{2017MNRAS.469.2806A} analysed the spatially resolved MS in a sample of 93 local massive galaxies on $\sim$ kpc scales. They find that the slope of the relation varies dramatically (from 0.3 to 0.99) depending on the range of $\Sigma_{\star}$ used for fitting, somehow recalling the bending of the MS observed in the integrated MS relation \citep[e.g.][]{2019MNRAS.483.3213P}. In addition, they find a scatter that varies between 0.55 and 0.7 and that they ascribe to a combination of local variations of the specific SFR (sSFR) and different large scale galaxy properties (i.e. morphology, existence of a bar). \citet{2019ApJ...887..204J} found a spatially resolved MS in galaxies at 0.1 < $z$ < 0.42 observed with MUSE. \citet{2018MNRAS.475.5194M} estimated the MS from $\sim$ 800 galaxies in the Sydney AAO Multi-object Integral Field Galaxy Survey \citep[SAMI,][]{2012MNRAS.421..872C} at $z < 0.1$, finding a slope of 1.0. \citet{2018ApJ...865..154H} studied the spatially resolved MS in a sample of 93 nearby galaxies drawn from the Survey for Ionized Neutral Gas in Galaxies \citep[SINGG,][]{2006ApJS..165..307M} and the Wide-field Infrared Survey Explorer \citep[WISE,][]{2010AJ....140.1868W} on scales ranging from 50pc to 10kpc. They find that the slope, scatter and normalisation of the relation do not vary when changing the spatial scale, up to 1-1.5kpc. They also find no dependence of the scatter (of both the spatially resolved and the global relation) on structural parameters of galaxies (morphology) and their HI gas fraction \citep[as previously suggested for the global MS by, e.g.,][]{2016MNRAS.462.1749S}. Thus, they conclude that the MS scatter can be driven by systematic differences in the star formation process among galaxies or their global environment. \citet{2019MNRAS.488.3929C} exploited a sample of 2000 MaNGA galaxies to study the global and spatially resolved MS and its dependence on morphology. They find a spatially resolved MS with a slope close to 1 and a scatter of 0.27 dex, around which SF areas are located depending on the galaxy integrated morphology. They conclude that some processes related to morphology must set the local SF activity of a galaxy. Recently, \citet{2019MNRAS.488.1597V} studied the spatially resolved MS in a sample of 30 local late-type star-forming galaxies up to four effective radii, exploiting the GAs Stripping Phenomena in galaxies with MUSE survey \cite[GASP,][]{2017ApJ...844...48P}. They find that $\Sigma_{\star}$ and $\Sigma_{\rm SFR}$ correlate with a scatter of 0.3 dex, larger than the one observed for CALIFA and MaNGA galaxies. Interestingly, this correlation is found to vary dramatically when studied for single galaxies: for several sources a correlation is not even present, for others multisequences are observed, undermining the existence of a universal correlation. 

To reconcile and homogenize all these observational results into a coherent scenario, it is mandatory to account for the different selection samples and for potential systematics in the methodology adopted to compute the galaxy physical parameters. To this aim, we try to cope with the main biases affecting the bulk of the previous works that are almost based on emission lines (i.e. H$\alpha$) or UV-to-optical tracers to constrain the SFR (see Sanchez et al., 2019 for a review). Here, we exploit the power of a complete multiwavelength photometric coverage, extending from the far-UV up to the far-IR (i.e. from {\em GALEX} to {\em Herschel}) to provide a reliable measure of the comprehensive SFR in galaxies, directly accounting for both the observed and the obscured components. This is obtained by performing an SED fitting procedure, that, accounting for an energetic balance, allows to go deeper than spectroscopic surveys based on H$\alpha$ emission lines. The latters critically depend on dust extinction corrections derived from the Balmer decrement and mostly  on the spectral Signal-to-Noise ratio (SNR) limit needed (for example) for the spaxels classification on BPT diagrams, requiring significant detection of four emission lines. Thus, we exploit the higher SNR of the photometric data that allows to reach the farthest galactocentric distances, and the combination of UV and far-Infrared measurements to minimize the uncertainties related to the presence of dust.

With such a machinery, in this work (Paper I) we aim at defining the MS at kpc-scales in a sample of nearby face-on spirals collected in DustPedia \citep{2017PASP..129d4102D}. This choice is driven by the need to firstly characterize the final products of pure secular evolution processes, i.e. unperturbed disks, that have assembled at earlier cosmic epochs the bulk of the stars that we observe in today's galaxies \citep{2011ApJ...739L..40R, 2012ApJ...747L..31S}. Indeed, grand-design spiral galaxies are considered to account for the largest percentage of star forming galaxies sitting on the local MS (Wuyts et al. 2011, Morselli et al. 2017). This work focuses first on the study of typical MS systems and it will be extended in the future to other morphological type galaxies, that are statistically located at larger distances from the local relation. To reach a comprehensive picture on the star formation process in galaxies, in a companion paper (Paper II, Morselli et al. in prep.) we will investigate possible variations of the star formation efficiency (SFE) within galaxies, as well as the relation between cold gas availability and SFR.

The paper is structured as follows. In Section 2 we describe the dataset. In Section 3 we present our SED fitting methodology (e.g. libraries, image processing). In Section 4 we present our results on the spatially resolved MS. Thorough the paper we will assume a $\Lambda$CDM cosmology \citep{2016A&A...594A..13P} and \citet{2003PASP..115..763C} IMF.

\section{Data and Sample}\label{sec:Data}

We exploit the vast amount of data from DustPedia\footnote{The DustPedia website is available at http://dustpedia.astro.noa.gr}, a collection of more than 875 local galaxies, built with the purpose of studying the dust emission in the local universe. DustPedia comprises every {\it Herschel} observed galaxy within $\sim$ 50 Mpc from the Milky Way, with a diameter $D_{25} > 1^{\prime}$. The multiwavelength photometric data presented in DustPedia have been homogenised, making this dataset particularly interesting for our purposes. 
\begin{figure*}
	\includegraphics[width=\textwidth]{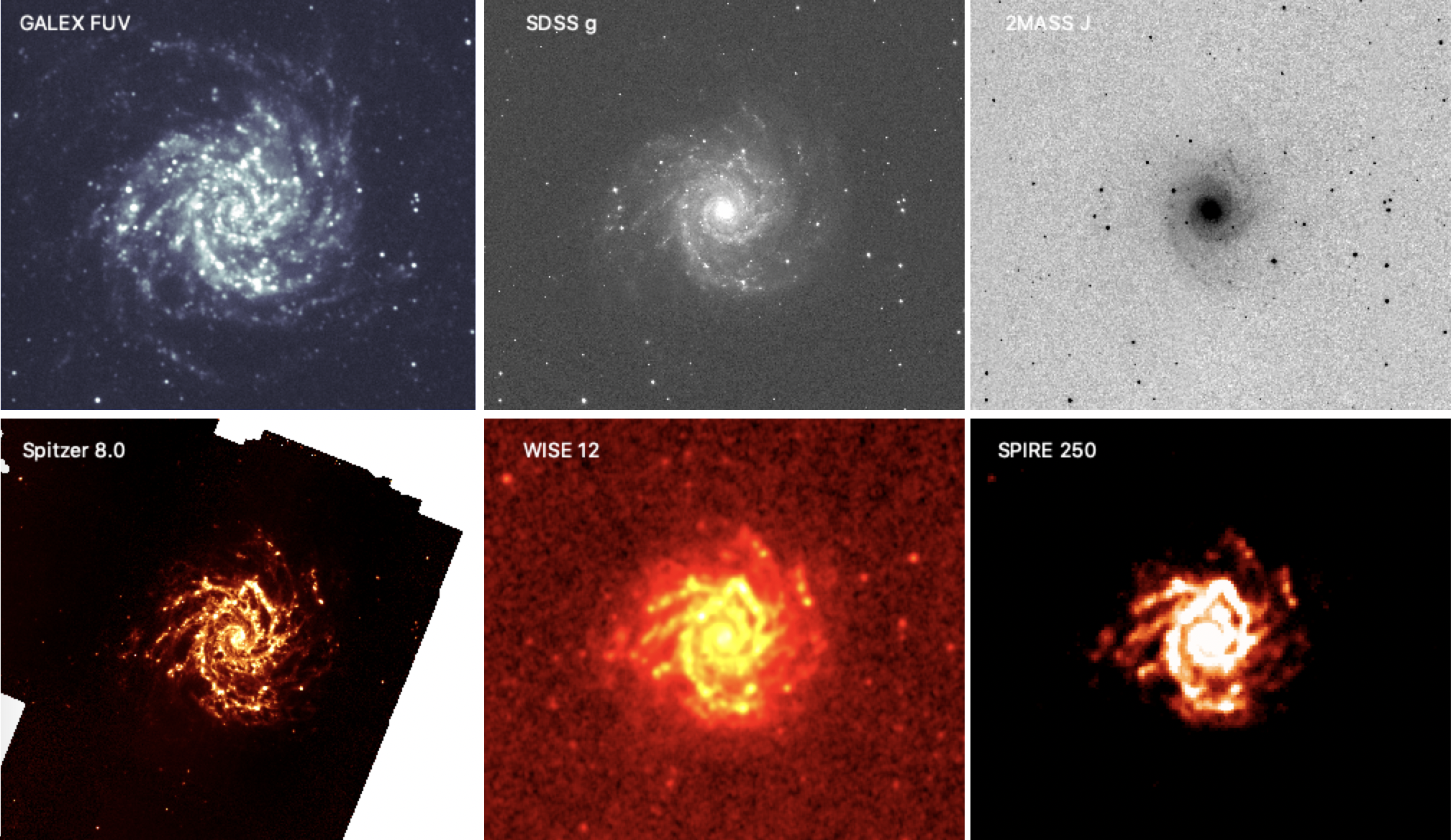}
    \caption{
    Example of the multiwavelength imaging available for NGC0628, coming from the DustPedia archive. Image size is $\sim 8^{\prime} \times 8^{\prime}$.
    }
    \label{fig:ex_bands}
\end{figure*}

\subsection{Sample selection}

Starting from the DustPedia collection, we build a coherent sample of local star forming galaxies with a "grand-design spiral" structure. Thus, we select galaxies having a Hubble stage index T \citep[or RC3 type,][]{1991rc3..book.....D, 1994AJ....108.2128C} between 2 and 8, and a diameter $D_{25} > 6'$. To avoid complications related to disk inclination ({\it i}) and dust correction, we focus here on nearly face-on sources with {\it i} $< 40^{\circ}$. Furthermore, we apply a cutoff in distance (around 1000 km/s, corresponding to approximately 22 Mpc), as for galaxies beyond this distance the sub-mm photometry is scarcely resolved or even below the observational beam. Finally, to robustly estimate the physical parameters of galaxies from SED fitting, we restrict our analysis to those having a uniform coverage in the wavelength domain, with at least 20 bands of observation. 

The final sample includes 8 objects: NGC0628, NGC3184, NGC3938, NGC4254, NGC4321, NGC4535, NGC5194, NGC5457. Distances, sizes, inclinations and morphologies are those adopted by the DustPedia collaboration, based on the HyperLEDA database\footnote{The HyperLEDA database is available at http://leda.univ-lyon1.fr/} \citep{2014A&A...570A..13M}, and are available on their website (see Tab.\,\ref{tab:sample}). The sample stellar masses are in the range between $10^{10}-10^{11} M_{\odot}$, and star formation rates between 1 and 4 $M_{\odot}$ yr$^{-1}$ These sources represent the evolved descendants of high-redshift normal star-forming galaxies, that are rotationally supported systems, at least by $z=2$ \citep[e.g.][]{2015ApJ...799..209W, 2009ApJ...706.1364F, 2018ApJS..238...21F}. Moreover, their regular dynamical properties suggest that they have quiescently assembled their stellar mass through cosmic times. In Fig.\,\ref{fig:ex_bands} we present, as an example, the galaxy NGC0628 as it is observed in the far UV (FUV) with the GALaxy Evolution eXplorer \citep[GALEX,][]{2007ApJS..173..682M}, in the g-band with the Sloan Digital Sky Survey \citep[SDSS,][]{2000AJ....120.1579Y}, in the J-band with the 2 Micron All-Sky Survey \citep[2MASS,][]{2006AJ....131.1163S}, at 8$\mu$m with the {\it Spitzer} Space Telescope \citep{2004ApJS..154....1W}, at 12$\mu$m with WISE and finally at 250$\mu$m with the {\it Herschel} Space Observatory \citep{2010A&A...518L...1P}. In the following paragraphs we briefly report information on the different data sets used in this work. For a more systematic and complete reference about data reduction and processing (implemented by the DustPedia collaboration) we refer the reader to \citet{2018A&A...609A..37C}.

\begin{table*}
	\centering
	\caption{Our galaxy sample. Galaxy name, coordinates in J2000 system reference, distances D (in Mpc), inclinations, r$_{25}$ sizes (the semimajor axis isophote at which the optical surface brightness falls beneath 25 mag arcsec$^{−2}$) and morphological classifications are the same adopted by the DustPedia collaboration, and come from the HyperLEDA database \citep{2014A&A...570A..13M}. The values of M$_{\star}$ and SFR are obtained fitting the DustPedia photometry with {\sc magphys}, the former as the standard {\sc magphys} output, the latter from the scaling relations in Eq.\,\ref{eq:Luv-SFR_scalrel}-\ref{eq:LIR-SFR_scalrel}. It is worth noticing that NGC5194 aperture photometry comprehends both fluxes coming from M51a and M51b. Reported cell sizes are referred to the so-called {\it pixel-by-pixel} SED fitting (in this case, with 8$\arcsec$ square size, reported in kpc) and regions probing a fixed physical scale (1.5 kpc, in arcseconds), as reported in Sec.\,\ref{sec:Methods}.}
	\label{tab:sample}
	\begin{tabular}{lccccccccccc}
		\hline
		Galaxy Name     & RA & DEC & D & i & r$_{25}$ & $\log$M$_{\star}$ & SFR & \multicolumn{2}{c}{Cell size}  & RC3 Type        \\
		                & [deg] & [deg] & [Mpc] & $[\circ]$ & kpc & $[M_{\odot}]$ & $[M_{\odot}$/yr]    & 8$\arcsec$ [kpc] & 1.5 kpc [$\arcsec$] &           \\
		\hline
    	NGC 0628 (M74)  &  24.1740 & 15.7833 & 10.14 & 19.8 & 14.74 & 10.41$\pm$0.15 & 1.90$\pm$0.41 & 0.39 & 30.60 & Sc  \\
        NGC 3184        & 154.5708 & 41.4244 & 11.64 & 14.4 & 12.55 & 10.14$\pm$0.10 & 0.98$\pm$0.10 & 0.45 & 26.67 & SABc \\
        NGC 3938        & 178.2057 & 44.1208 & 19.41 & 14.1 & 10.04 & 10.16$\pm$0.20 & 2.19$\pm$0.19 & 0.75 & 16.03 & Sc \\
        NGC 4254 (M99)  & 184.7065 & 14.4164 & 12.88 & 20.1 &  9.40 & 10.02$\pm$0.18 & 2.44$\pm$0.23 & 0.50 & 24.11 & Sc  \\
        NGC 4321        & 185.7282 & 15.8219 & 15.92 & 23.4 & 14.30 & 10.74$\pm$0.15 & 3.27$\pm$0.37 & 0.62 & 19.52 & SABb \\
        NGC 4535        & 188.5845 &  8.1978 & 14.93 & 23.8 & 17.62 & 10.19$\pm$0.19 & 1.30$\pm$0.08 & 0.58 & 20.81 & Sc  \\
        NGC 5194 (M51)  & 202.4695 & 47.1952 &  8.59 & 32.6 & 17.23 & 10.70$\pm$0.20 & 4.08$\pm$0.26 & 0.33 & 36.10 & Sbc  \\
        NGC 5457 (M101) & 210.8025 & 54.3491 &  7.11 & 16.1 & 24.81 & 10.38$\pm$0.13 & 2.48$\pm$0.15 & 0.28 & 43.60 & SABc \\

		\hline
	\end{tabular}
\end{table*}

\subsection{GALEX}
The ultraviolet part of the electromagnetic spectrum is sampled by GALEX. GALEX data are divided in near-UV (NUV, 1516 \AA, with full-width at half maximun (FWHM) of $\sim 4.25\arcsec$) and FUV (2267 \AA\,and $\sim 5.25\arcsec$ beam). NUV and FUV data are available for every galaxy in our sample, and sample the light coming from newborn massive stars, tracing the unobscured star formation activity of galaxies.

\subsection{SDSS}
The optical part of the spectrum, sampling the young stellar content, is probed with the Data Release 12 of SDSS. We use five bands ($u$, $g$, $r$, $i$, $z$), with effective wavelengths of 3351, 4686, 6166, 7480 and 8932 \AA. The SDSS FWHM varies from observation to observation, with a typical median value of 1.43$\arcsec$.

\subsection{2MASS}
Near-infrared (NIR) imaging comes from 2MASS in three bands, $J$-band (1.25 $\mu$m), $H$-band (1.65 $\mu$m) and $K_{s}$-band (2.16 $\mu$m), with FWHMs between $\sim 2.5\arcsec$ and $2.7\arcsec$. As noted in Appendix D of \citet{2018A&A...609A..37C}, 2MASS-H band photometry might be affected by a particular offset with respect to $J$ and $K_s$ band photometry, translating as a "bump" or "dip" in the SED of the object. We find this problem only NGC0628 photometry, for which we exclude the H-band data-point from the SED fitting.

\subsection{WISE and Spitzer}
The NIR and medium-infrared (MIR) parts of the spectrum are observed with the WISE and the IRAC camera on board of {\it Spitzer}. WISE provides observations at 3.4, 4.6, 12 and 22$\mu$m, with Point Spread Function (PSF) FWHMs ranging from 5.70$\arcsec$ to 11.8$\arcsec$. The IRAC camera, instead, collects observations at 3.56, 4.51, 5.76 and 8.00$\mu$m, with a PSF FWHM varying between 1.66$\arcsec$ and 1.98$\arcsec$. These bands are available for the whole sample, except for NGC4535 and NGC5457, both lacking 5.8$\mu$m data. NIR and MIR observations trace the old stellar component, the stellar mass distribution and the carbonaceous-to-silicate materials in the dust.

\subsection{Herschel}
Emission from the far infrared (FIR) to the sub-mm is observed with the instruments on board the {\it Herschel} Space Observatory. We used both {\it Herschel} PACS (70$\mu$m, 100$\mu$m and 160$\mu$m) and SPIRE data (250$\mu$m and 350$\mu$m), with PSF FWHM of $\sim$6, 8, 12, 18 and 24$\arcsec$ respectively. The only exceptions are NGC4535 and NGC5194, lacking 70$\mu$m and 100$\mu$m observations, respectively. These wavelengths typically probe the reprocessed emission coming from  dust, and thus could constrain the dust-obscured star-formation processes. We do not include the 500$\mu$m channel in our SED fitting procedure. Indeed, in our approach (see Sec. \ref{sec:Methods}) we need to downgrade all images to the lower resolution available in our maps. With a beam of $\sim35\arcsec$, the 500$\mu$m does not allow to push our spectro-photometric analysis to the kpc scale. However, we have verified that the performed SED best-fits are consistent with the observed 500$\mu$m flux densities. The adopted spatial scale for our analysis will be discussed in details in Sec.\,\ref{sec:Methods}.

\section{Methodology and SED fitting}\label{sec:Methods}
To obtain the spatially resolved physical properties of galaxies, like their stellar mass (M$_{\star}$) and SFRs, we perform a SED fitting on scales varying from 0.28 kpc to 1.5 kpc.

Our procedure includes three steps: (i) matching the PSF of every image to the worst-one, by degrading every band to the PSF of the SPIRE 350 image; (ii) building a grid of square cells of a given size and measuring the flux at each wavelength on them; (iii) deriving the physical properties of the individual cells by performing SED fitting to the available photometry.

We analyse each galaxy at two different resolutions: the first by considering cells of $8\arcsec \times 8\arcsec$ (thus a varying length side in physical scale from one galaxy to another, as reported in Tab.\ref{tab:sample}), that is the pixel scale of SPIRE 350 maps, corresponding to three spaxels per resolution element, and the second by constructing fixed size cells of 1.5 kpc $\times$ 1.5 kpc, as this is the typical physical scale sampled in most of the literature. This is done to compare the reliability of the SED fitting procedure and the final results obtained sampling different spatial scales. 

Finally, we also perform SED fitting on the integrated galaxy photometry coming from the DustPedia Photometric sample. This information is crucial to validate our photometric analysis (comparing with the public integrated properties of the DustPedia sample), and to understand the regulation of star formation properties on galactic scales as compared to sub-galactic scale, that is the main aim of this work. We also checked that the DustPedia integrated photometry is compatible with the one we obtain within the same apertures.

\subsection{Image preprocessing and PSF degradation}
As reported in Sec.\,\ref{sec:Data}, the starting point are the different photometric observations of the galaxies, downloaded from the DustPedia Archive. These have already been homogenised in flux, given in Jy, and World Coordinate System. For our purposes, on each map we perform: foreground stars removal, background estimation, flattening and subtraction, and PSF degradation, matching the maps resolution to the worst one.

Foreground stars removal is performed exploiting the Comprehensive \& Adaptable Aperture Photometry Routine (CAAPR) routine presented in \cite{2018A&A...609A..37C}. The routine uses the PTS toolkit for SKIRT \citep{2015A&C.....9...20C} to detect and remove/patch foreground star emission, thus creating a star subtracted version of the map. As CAAPR sometimes mistakenly identifies bright HII regions as stars, we carefully visual-check the star-subtracted maps to be sure that only foreground stars are removed. 

In each photometric band, we perform background estimation and subtraction on the star-subtracted maps. This procedure follows the indications in \cite{2018A&A...609A..37C}, performing (if needed) a sky flattening fitting with a 5th order polynomial 2D array, and then subtracting the background emission. This step is crucial to remove galactic foreground emission and to smooth out residual image gradients typical of some bands (i.e. GALEX, Spitzer).

Finally, we degrade the stars and background subtracted maps to the PSF of SPIRE 350. To do this, we convolve the maps using the kernels provided by \cite{2011PASP..123.1218A}\footnote{http://www.astro.princeton.edu/$\sim$ganiano/Kernels.html}, in order to match each band to the PSF of SPIRE 350.

\subsection{Spatially resolved SED fitting}
\label{subsec:sed}
To perform SED fitting, we measure the flux in each photometric band inside the cells with the {\sc photutils v0.6} Python package \citep{Bradley_2019_2533376}. Then, we correct all the fluxes with wavelength lower than 10 $\mu$m for Galactic extinction, with the in-built module in CAAPR, based on values in the IRSA Galactic Dust Reddening and Extinction Service. To estimate errors on the flux, we apply the following procedure. When available, we use the error maps in the DustPedia Archive (i.e. Spitzer bands and the far-IR photometry). If no error map is available (i.e. the SDSS maps), we take the signal-to-noise ratio of the DustPedia Photometry in that particular band (accounting also for calibration error), and use that to evaluate the error in each cell (this is different from what done by \citet{2018MNRAS.476.1705S}, that adopt a default SNR of 5 for every band). As a result we obtain two photometric catalogues (for the $8 \arcsec \times 8 \arcsec$ cell and for the 1.5kpc $\times$ 1.5kpc cell) containing the flux in every available photometric band from UV to far-IR, the relative errors, and the associated astrometric positions of the pixel/cell centres. We include in the SED fitting procedure only cells with no more than 5 undetected bands. The discarded cells are mostly found in the outermost part of galaxies. This cut critically reduces the total computational time of the subsequent SED fitting procedure. 

To perform SED fitting, we use the publicly available code {\sc magphys} \citep{2008MNRAS.388.1595D}. {\sc magphys} is one of the state-of-art code to model panchromatic SED, and as such has been extensively used in the literature \citep[e.g.][]{2010A&A...523A..78D, 2012MNRAS.427..703S, 2013A&A...551A.100B, 2015MNRAS.446.1512H, 2015MNRAS.453.1597S, 2015ApJS..219....8C, 2018MNRAS.475.2891D, 2018MNRAS.478....2C, 2018MNRAS.479..297W, 2019ApJ...882...65M}. It simultaneously models the emission observed in the UV-to-FIR regime by consistently assuming that the whole energy output is balanced between the one emitted at UV/optical/NIR wavelengths that is absorbed by dust and the one re-emitted in the FIR. It does so exploiting a Bayesian approach to determine the posterior distribution functions of the parameters by fitting the observed photometry to the model emission coming from a set of native libraries within the {\sc magphys} distribution. These libraries are composed of 50000 stellar population spectra with varying star forming histories (described by a continuous model $\psi (t) \propto \exp^{- \gamma t}$ with superimposed random bursts) derived from the \citet{2003MNRAS.344.1000B} spectral library. Formation ages are uniformly distributed between 0.1 and 13.5 Gyr (thus they are shorter or equal to the age of the Universe for the nearby galaxies in this work). The typical SF timescales are uniform between 0 and 0.6 Gyr$^{-1}$, dropping exponentially around 1 Gyr$^{-1}$. Metallicities are taken from a uniform grid with values ranging between 0.02 and 2 $Z_{\odot}$. The stellar emission libraries are associated, via the energy balance criterion, to 50000 two-components dust emission SEDs \citep{2000ApJ...539..718C, 2008MNRAS.388.1595D}, accounting for the emission coming from the so-called stellar birth clouds and the diffuse interstellar medium. {\sc magphys} has already been used to model panchromatic SED of local galaxies \citep[e.g.][for M31]{2014A&A...567A..71V} and its ability to properly retrieve physical properties from the SED has been tested on hydrodynamical simulations of galaxies \citep[i.e. isolated disk or major merger as in][]{2015MNRAS.446.1512H, 2015MNRAS.453.1597S} and of resolved regions \citep{2018MNRAS.476.1705S}. In particular, \cite{2018MNRAS.476.1705S} run {\sc magphys} on 21 bands of a simulated isolated local disk galaxy, on regions spanning from 0.2 to 10 kpc, also with different disk inclinations, in order to test its ability to recover physical properties (such as M$_{\star}$, SFRs, SFHs) and derived ones (i.e. metallicity, visual extinction $A_V$). They find that {\sc magphys} is able to produce acceptable fits to almost every cell enclosed inside the $r$-band effective radius, and between 59\%-77\% of cells within 20 kpc from the center (even though the SNR choice of 5 might play an important role in keeping the $\chi^2$ total value below the threshold). Those fits lead to reasonably good parameters estimate for every built-in {\sc magphys} output. 

Following \cite{2018MNRAS.476.1705S}, we input the measured fluxes to {\sc magphys} for SED fitting. As anticipated in Sec.\,\ref{sec:Methods}, we perform spatially resolved SED fitting for each galaxy using two different dimensions of the cell: 8$\arcsec$ and 1.5kpc. This is done for two reasons: $i)$ to compare the results on different spatial scales, and $ii)$ to test the reliability of the pixel-by-pixel SED analysis. In fact, the physical scales probed in this work vary from $\sim 280$ pc for NGC5457 (the closest galaxy in our sample) to $\sim 750$ pc for NGC3938 (the most distant, see Tab.\,\ref{tab:sample}). These scales are at the limit of the assumption on which the balance between star formation laws and energy output still holds \citep[e.g.][]{2003ApJ...590L...1K, 2017MNRAS.468..920K, 2010ApJ...722.1699S}. In Sec.\,\ref{sec:Results} we show how the 8$\arcsec$ results and the 1.5kpc ones are consistent, thus concluding that our analysis is reliable also on the smallest scales. 

\begin{figure*}
	\includegraphics[width=\textwidth]{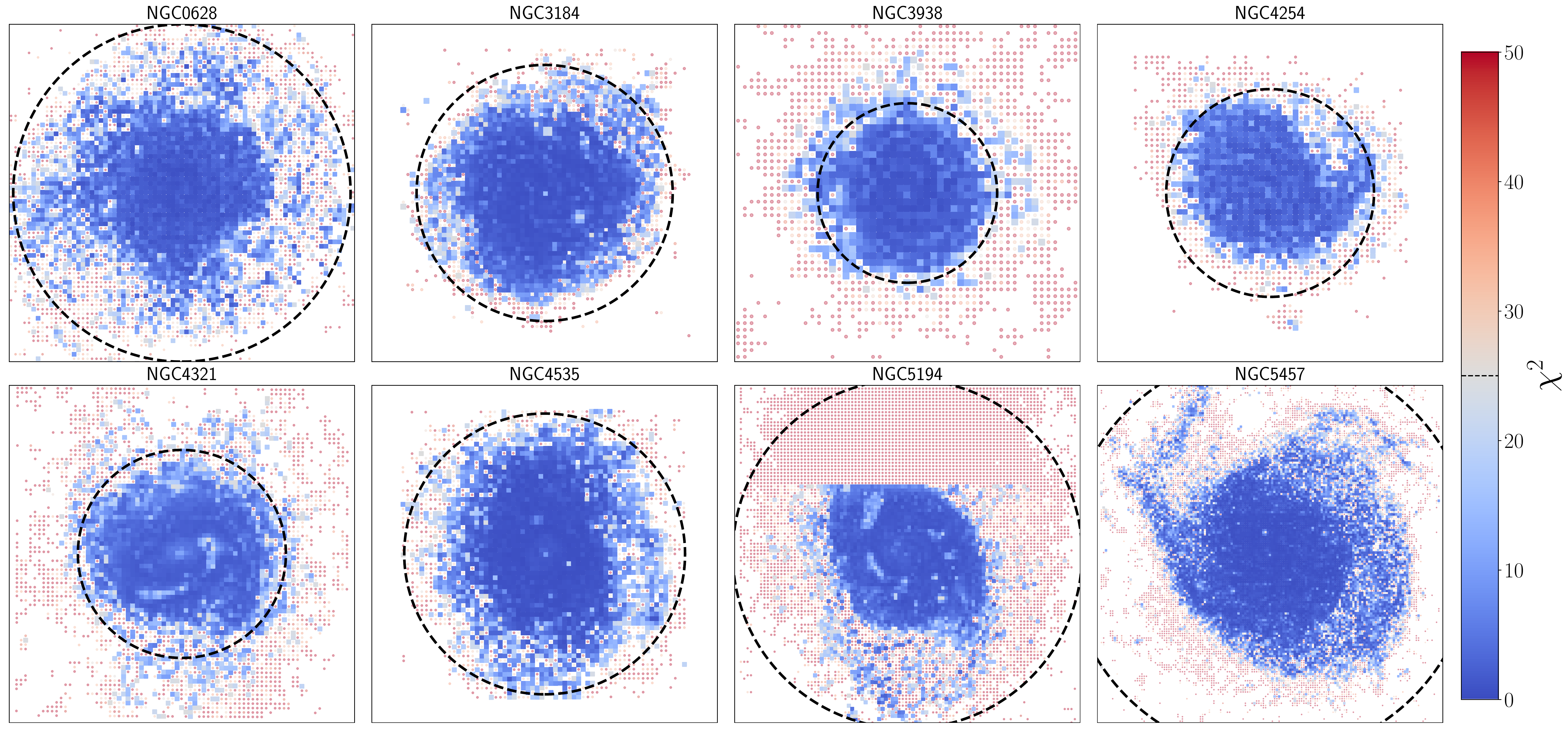}
    \caption{Final $\chi^2$ maps for the full sample of galaxies for cells of 8$\arcsec$. Blue colored squares are the accepted values (below the threshold of 25, highlighted with a dotted line in the colorbar), red shaded points the discarded ones lying over the threshold. For NGC5194, each point with $\delta > 47.24052$ has been discarded in order to remove contamination from the M51b companion (they form a second tight sequence below the accepted cells). The dotted circle is the galaxy $r_{25}$ reported in Tab.\,\ref{tab:sample}. Almost every cell in the innermost galaxy regions are accepted. In the outer part of the disk, the spiral arms tend to produce acceptable fits with respect to the non-spiral part of the galaxies, as it is clear from NGC5457.}
    \label{fig:chi_squares}
\end{figure*}
To accept or reject a fit, we apply a criterion based on \cite{2015MNRAS.446.1512H}, where they used a $\chi^2$ threshold of 20 on 17 simulated photometric bands. Differently, in \cite{2018MNRAS.476.1705S} the threshold is set to 30.6 on 21 simulated bands. As the galaxies in our sample have been observed with 21 (NGC4535) or 23 photometric bands, we use a conservative $\chi^2$ cut of 25. $\chi^2$ maps for each galaxy are reported in Fig.\,\ref{fig:chi_squares}. Blue squares are the values below the threshold, while red points are the rejected ones. In NGC5194 the companion galaxy M51b has been removed by further rejecting each cell with a declination $\delta$ over 47.24052. On average, each galaxy has $\sim 4000$ accepted points, with the exceptions of the closest ones, NGC5194 ($\sim 10000$) and NGC5457 ($\sim 15000$).

\subsection{Stellar mass and SFR estimates}
The output of the SED fitting process is a wide range of physical properties. To probe the MS, the two fundamental quantities are the stellar mass and the star formation rate. M$_{\star}$ values are directly taken from the {\sc magphys} output. SFRs are obtained by summing unobscured (SFR$_{\rm UV}$) and obscured (SFR$_{\rm IR}$) contributions. SFR$_{\rm UV}$ is estimated using the relation of \citet{2001ApJ...548..681B}:
\begin{equation}
    {\rm SFR_{UV}} = 0.88\times10^{-28} L_{\nu},
    \label{eq:Luv-SFR_scalrel}
\end{equation}
where $L_{\nu}$ in erg s$^{-1}$ Hz$^{-1}$ is evaluated from the SED fit at 150 nm.\\
SFR$_{\rm IR}$ is computed using the relation of \cite{1998ApJ...498..541K}:
\begin{equation}
    {\rm SFR_{IR}} = 2.64\times10^{-44} L_{\rm IR},
    \label{eq:LIR-SFR_scalrel}
\end{equation}
where $L_{\rm IR}$ in erg s$^{-1}$ is evaluated from the SED fit between 8 and 1000 $\mu$m (rest-frame). In both cases, the relations have been rescaled in accordance with a \citet{2003PASP..115..763C} IMF.

\begin{figure}
	\includegraphics[width=0.9\columnwidth]{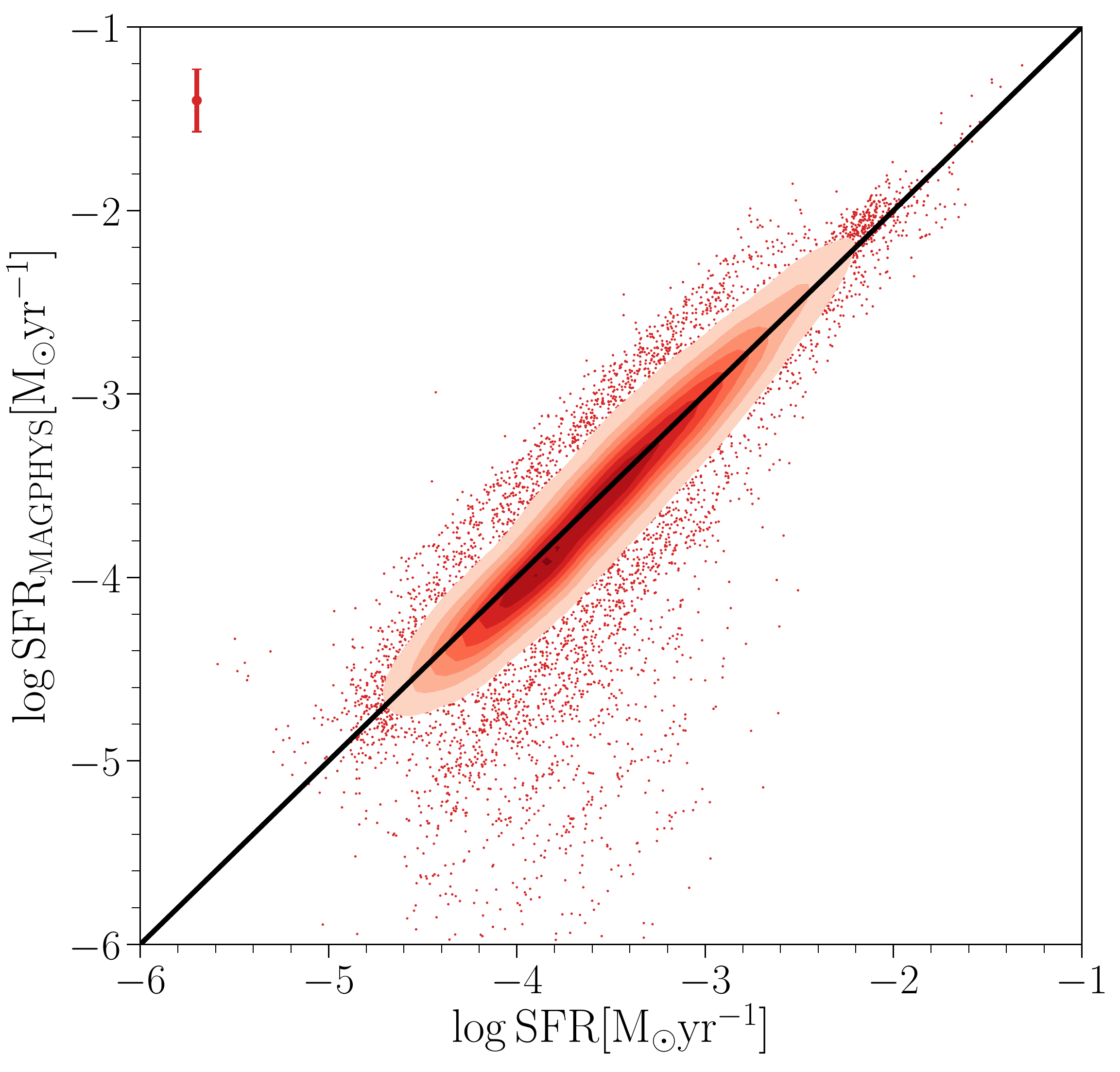}
    \caption{Comparison between the {\sc magphys} output star-formation rates (y-axis) and the ones we used thorough this work, obtained from the empirical relations in Eqs.\,\ref{eq:Luv-SFR_scalrel}-\ref{eq:LIR-SFR_scalrel} (x-axis). The red errorbar in the upper-left highlight the median error on the measure. Black line is the 1:1 relation. The vast majority of the sample is consistent with this relation, and only a small fraction (2.34\%) see their SFR increase by more than one order of magnitude.}
    \label{fig:SFR_comparison}
\end{figure}

We evaluate the SFRs from empirical relations that are less model dependent rather than using the SFRs obtained from the SED fit that are more dependent on degenerate parameters like age, metallicity, extinction. A comparison between the SFRs obtained from the empirical relations and the ones given as output of {\sc magphys} is shown in Fig.\ref{fig:SFR_comparison}. The scatter of the relation is relatively small, but the SFRs from {\sc magphys} are, on average, $\sim 0.1$ dex smaller than the ones from the empirical relations.
A detailed comparison of the output values of {\sc magphys} with those from other tools or receipts is beyond the goal of this paper. However, we note that the SED fitting technique accounts for the contribution of different stellar populations at various ages, as determined by the assumed SFHs. The empirical approach adopted here (and widely used in the literature) accounts for the contribution of two extreme populations: the UV emission from young stars, and the totally obscured young stellar component still hidden in the dusty molecular clouds. The tight correlation found in Fig.\,\ref{fig:SFR_comparison} ensures that the use of the SFR from {\sc magphys} would not change the results presented in the next Section. There are some points falling below the 1:1 relation, 369 with SFR increased by an order of magnitude, and 134 by two orders of magnitude. This is due to attributed star forming activity to what the {\sc magphys} models identify as (strong) diffuse cirrus emission in the spiral arms and in the inter-arms regions (the spatial distribution is Gaussian, centered on $0.5 r/r_{\rm 25}$). Anyway, we stress that the impact of these points to the overall results is non-existent, being an extremely small fraction (1.72\% and 0.62\% respectively) of the full sample.

\section{Results}\label{sec:Results}
\begin{figure*}
	\includegraphics[width=\columnwidth]{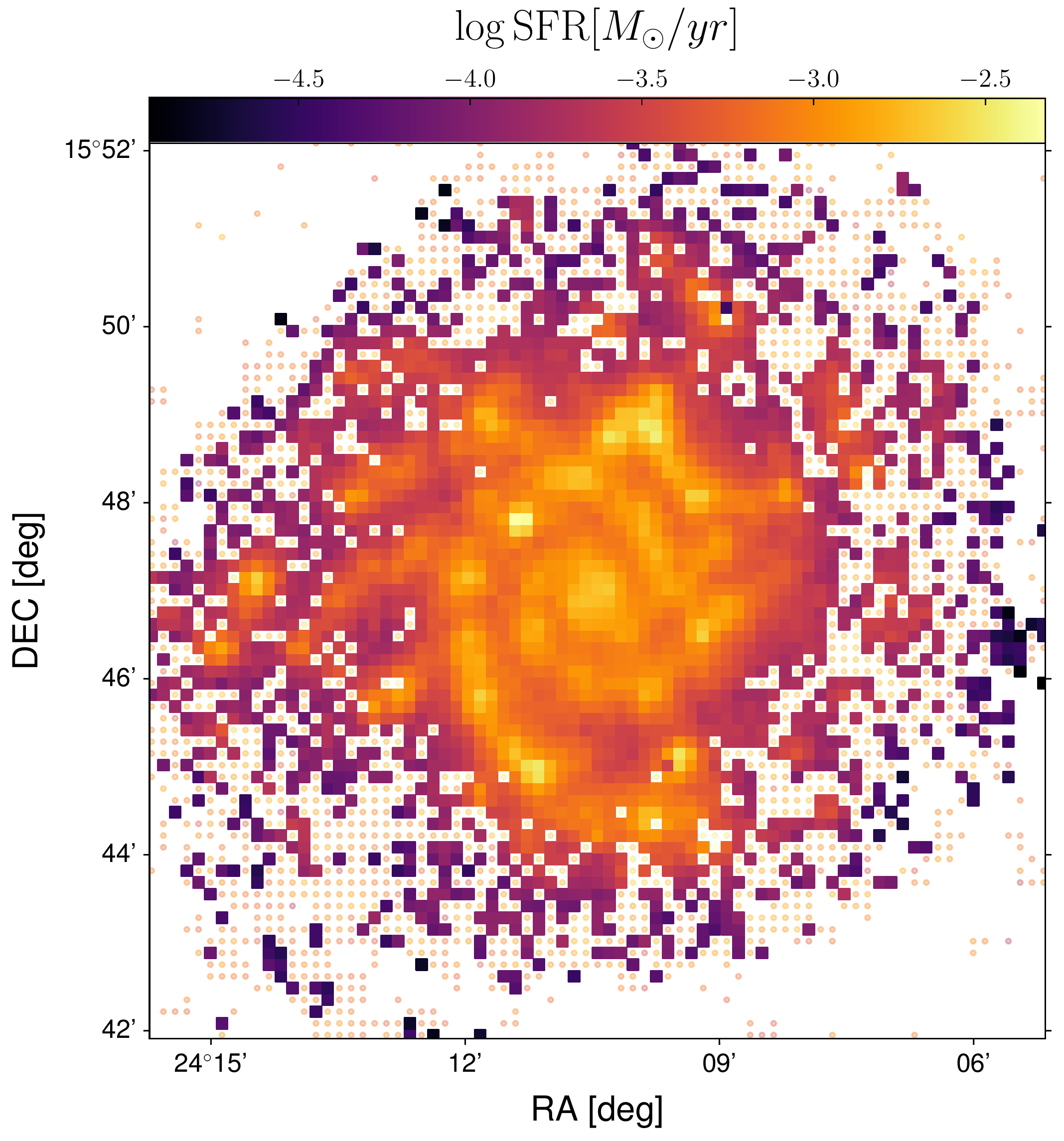}
	\includegraphics[width=\columnwidth]{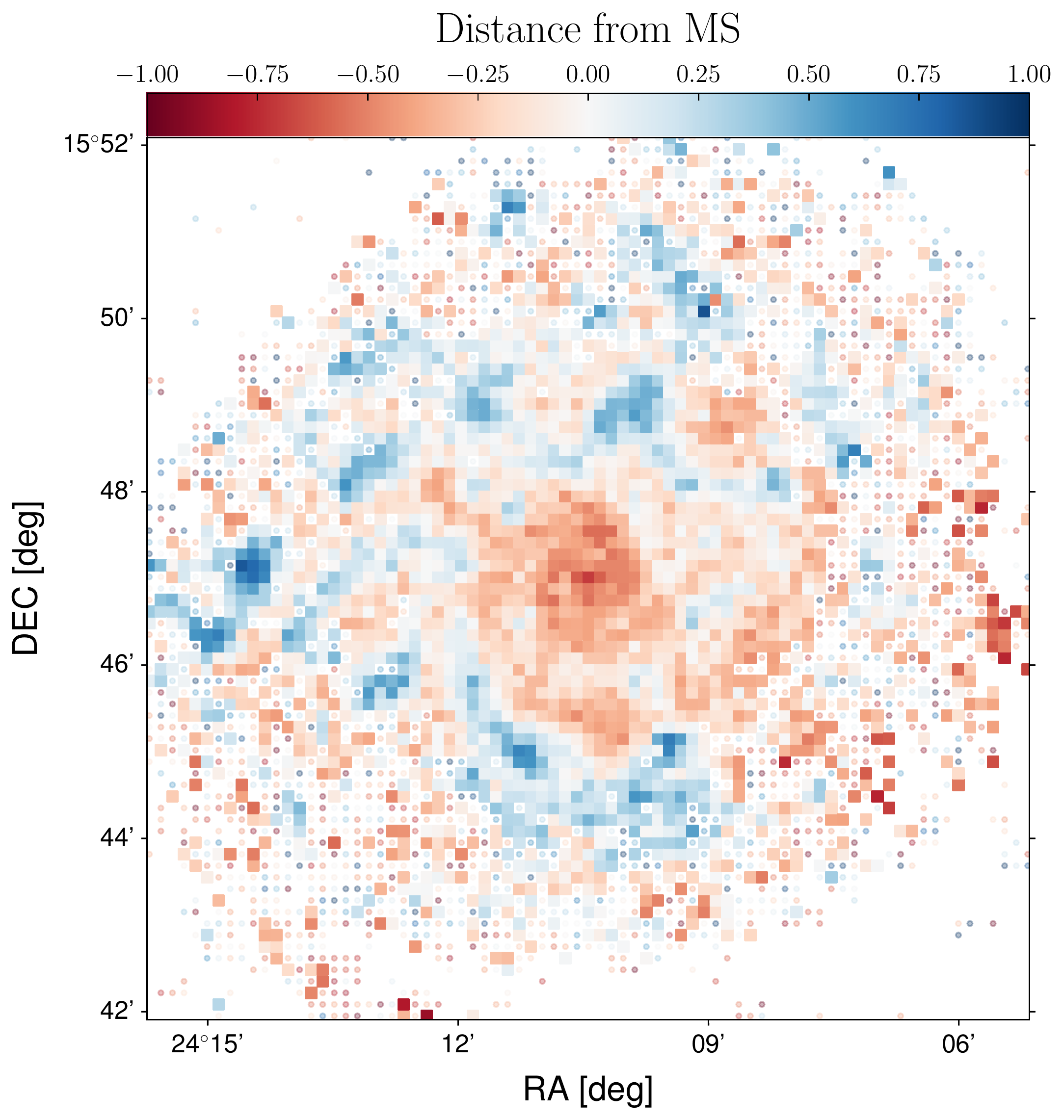}
	\includegraphics[width=\columnwidth]{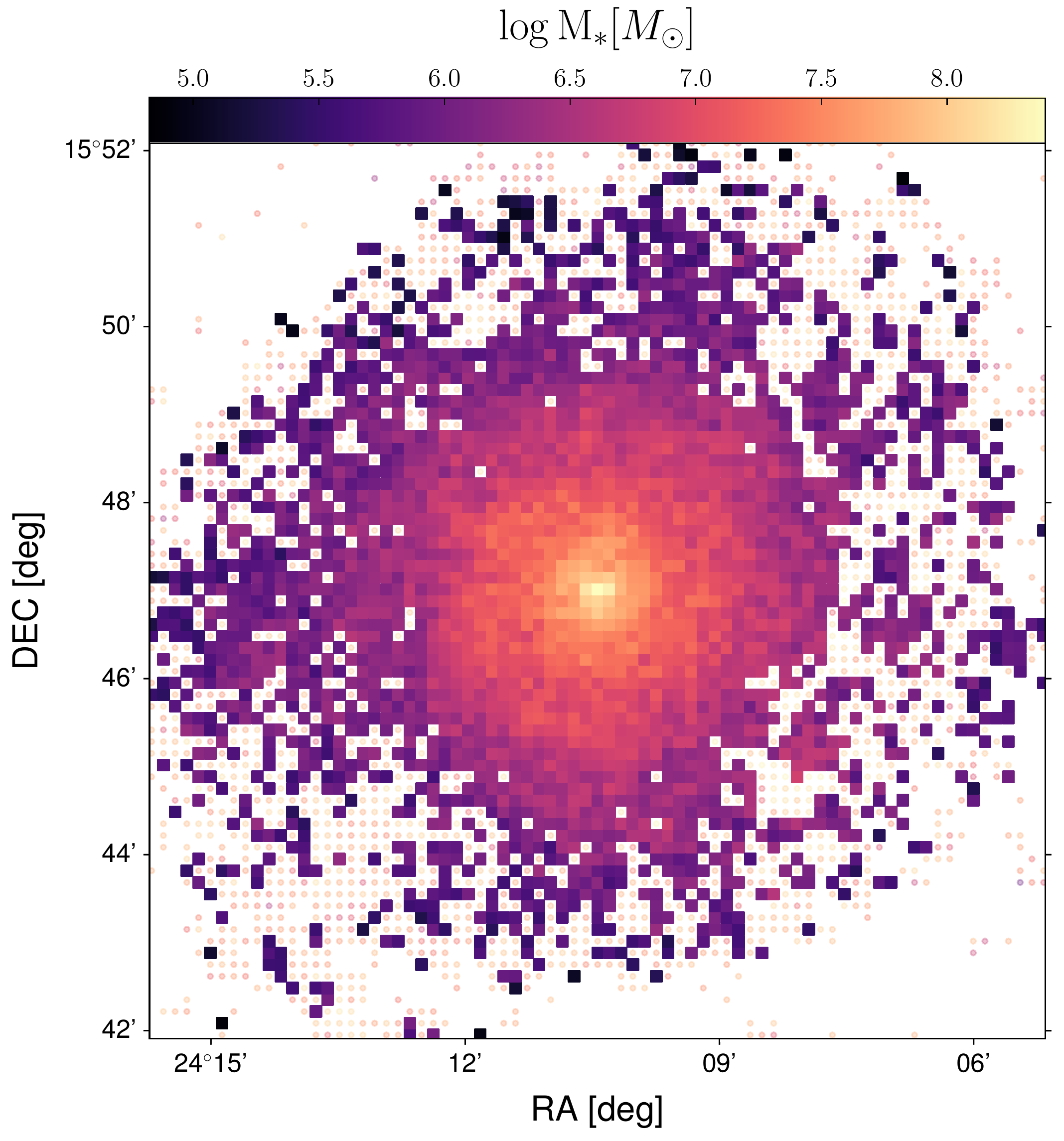}
	\includegraphics[width=\columnwidth]{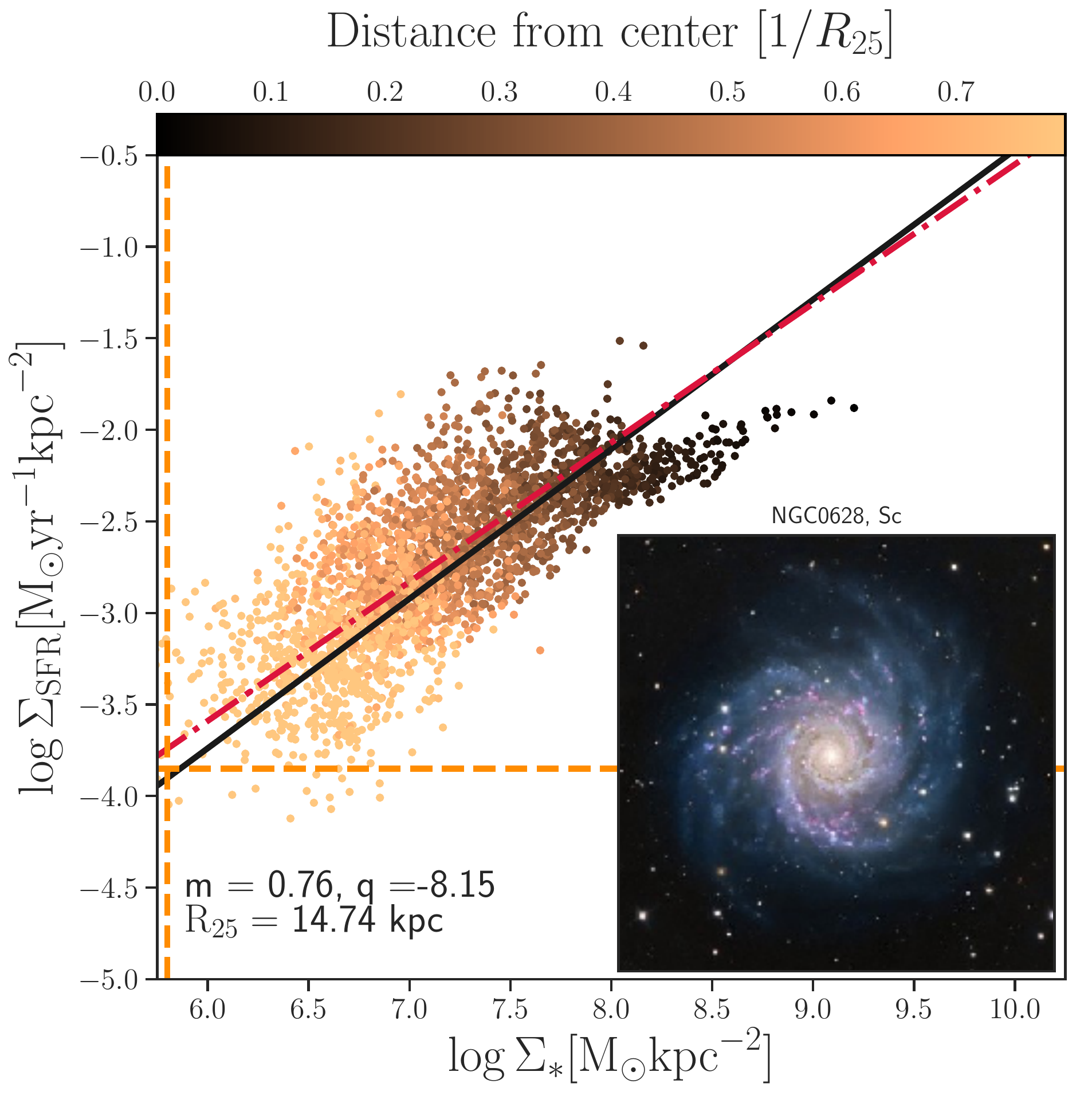}
    \caption{Summary plot for NGC0628. Panels are organized as follows: {\it upper left}, the cells $\log$ star formation rate; {\it upper right}, the cells distance from the MS defined in Sec.\,\ref{sec:Results}; {\it lower left}, the cells $\log$ stellar mass; {\it lower right}, how the galaxy cells are positioned in the $\Sigma_{\star}$-$\Sigma_{\rm SFR}$ plot, color-coded according to the distance from the galaxy center in units of $r_{25}$. Black line is the (total fitted) MS, red dashed line the ODR fit for the single galaxy points, orange lines the sensitivity limits, the inset an RGB image of the galaxy as observed in optical bands. In the first three panels, cells rejected for having a $\chi^2$ over the threshold are showed as points. Results for the rest of the sample are reported in Appendix\,\ref{App:results}.}
    \label{fig:NGC0628}
\end{figure*}
Before deriving any general conclusions and implications from the statistical analysis of the combined sample of local face-on spirals introduced in this work, we briefly present the main results about each individual source. An example for NGC0628 is shown in Fig.\,\ref{fig:NGC0628} (the others are in Appendix\,\ref{App:results}). In these Figures we present for each source: the maps of star formation rate, stellar mass and distance from the MS ($\Delta_{\rm MS}$ evaluated as the perpendicular distance of a point in the log$\Sigma_{\star}$-log$\Sigma_{\rm SFR}$ from the MS relation), using as reference the MS evaluated for the combined sample), as well as the distribution of the cells in the log$\Sigma_{\star}$ - log$\Sigma_{\rm SFR}$  plane, color coded as a function of the distance of the cell from the galaxy centre, in units of r$_{25}$. The dots in the maps correspond to rejected cells (with $\chi^2$ larger than 25), while the color-coded squares mark the cells with accepted $\chi^2$ (as explained in Sec.\,\ref{subsec:sed}). The black solid line in the bottom right panel of Fig.\,\ref{fig:NGC0628} and Figures A1-A7 is the MS relation of the combined sample, as explained in Sec.\,\ref{resolvedMS}.

It is evident from the maps that the SFR traces the spiral pattern of each galaxy (as seen in the RGB image in the inset of the lower right panel). In addition, we can see that for almost all the sources the SFR distribution has a peak at the centre, as well as several other peaks along the spiral arms. The stellar mass distribution is smoother and more centrally concentrated than as expected for an exponential mass distribution; in fact, according to the morphological classification of the galaxies as well as their RGB images, a bulge component is present in all our sources. The map of distance from the MS relation reveals that the spiral arms tend to lie above the relation for all the sources. The central bulge can be less (as in NGC0628, NGC3184, NGC3938, NGC5457) or more (as in NGC4254, NGC4321, NGC5194) starforming than the spiral arms. For those galaxies that have a 'red' bulge, a 'bending' of the MS relation is observed at the high $\Sigma_{\star}$ end, as the cells corresponding to the bulge component are also the more massive ones. It is worth noting that for the galaxies in our sample few spaxels would be considered passive (1 dex below the MS). This is a consequence of the sample selection, made of grand-design spirals. Also, due to the sample selection, there are no morphological trends in the log$\Sigma_{\star}$-log$\Sigma_{\rm SFR}$ plane, i.e. in scatter from the global (or individual) MS or in slope. This aspect will be the subject of a future work, where we plan to extend the same analysis to all kind of morphological types within DustPedia.

The two galaxies that are infalling in the Virgo cluster, NGC4254 and NGC4321, show two peculiar characteristics. NGC4254 is almost entirely made of cells that are located above the spatially resolved MS relation; indeed, this galaxy is the one that, according to its integrated properties, is located at the largest distance from the integrated MS relation (see Fig\,\ref{fig:GlobalProp}, X symbol), but still within the MS scatter. NGC4321 and NGC5194, instead, show the largest difference in slope with respect to the average MS relation. Interestingly, it also shows a second 'quenched' sequence below the MS, also observed in galaxies such as NGC4535 and NGC5194, and visible as a ring of red cells in Figures A4, A5 and A6.

\subsection{Spatially resolved MS}\label{resolvedMS}
\begin{figure}
	\includegraphics[width=\columnwidth]{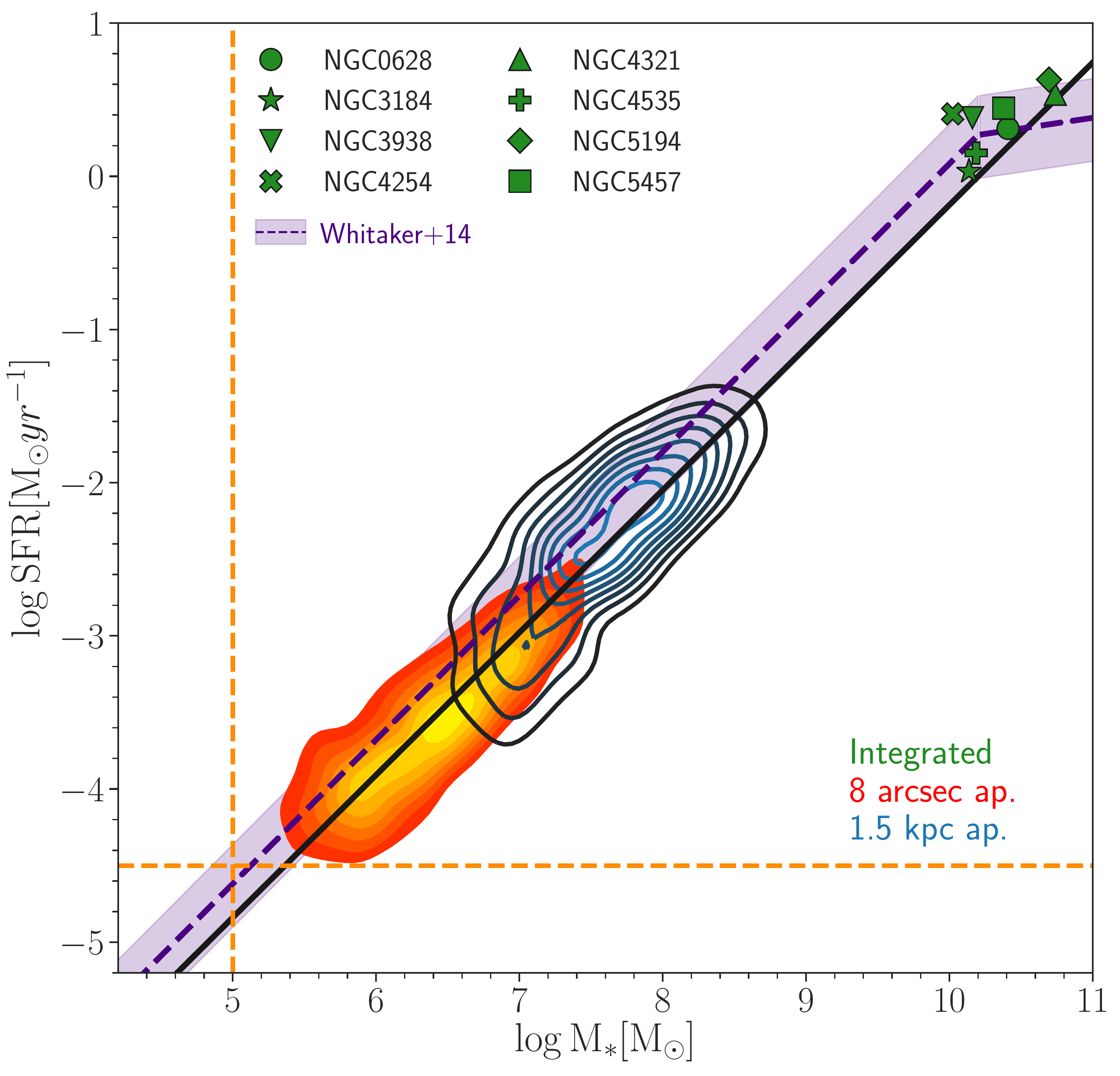}
    \caption{Sample properties evaluated in three different apertures. In green, the integrated properties, obtained fitting the available photometry in apertures containing the full galaxy emission. Marker size is as big as the data error bars. Blue and filled red contours are referred to cells of 1.5 kpc and 8$\arcsec$ respectively. Black line is the fit to the MS described by the 8$\arcsec$ results. Purple dashed line is the MS relation from \citet{2014ApJ...795..104W} rescaled at $z \sim 0$, while the shaded area enclose the scatter of MS distribution we obtain (0.27 dex). Dashed orange lines are the M$_{\star}$ and SFR sensitivity limit.}
    \label{fig:GlobalProp}
\end{figure}

In this Section, we analyse the spatially resolved MS using two approaches: 1) by studying the distribution of the spatially resolved parameters in the logM$_{\star}$ - logSFR plane, and 2) by analysing the distribution in the log$\Sigma_{\star}$ - log$\Sigma_{SFR}$ plane. The first approach gives us the possibility to understand whether the integrated MS relations, computed using SFR and M$_{\star}$ estimated over the full extent of galaxies, apply also at lower scales, thus hinting towards the existence of a universal relation at all scales. In Fig.\,\ref{fig:GlobalProp} we show the logM$_{\star}$ - logSFR plane and how the different cell in which we decompose the galaxies in our sample populate this plane. In particular, we show the results obtained using two different aperture sizes: 1.5 kpc (blue contours) and $8\arcsec$ (red-to-yellow gradient). We also plot, with green symbols, the integrated logM$_{\star}$ and logSFR values obtained via SED fitting on apertures that enclose the full galaxy. The orange dashed lines are the sensitivity thresholds, and correspond to the rms computed from the M$_{*}$ and SFR maps on regions far away from the galaxy emission. We have also checked that the rms limit for the SFR is consistent with the one obtained from the SPIRE maps. The red solid line is our linear fit obtained including only data points over the sensitivity thresholds (orange dashed lines); it has slope of 1.03 and intercept -10.17.
The contours referring to different spatial scales and the integrated quantities all lie along the the same relation. The logM$_{\star}$ - logSFR relation holds on a large variety of scales, from ones slightly over the size of the Giant Molecular Clouds \citep[GMCs, $\sim$ 200 pc,][]{1987ApJ...319..730S} to the galaxy as a whole. This results is corroborated by the recent work of \citet{2019arXiv191103479C}, showing that galactic star formation is driven by dynamical processes that are independent of the galaxy environment and are mostly governed by the GMC evolutionary cycling.

\begin{figure*}
	\includegraphics[width=\textwidth]{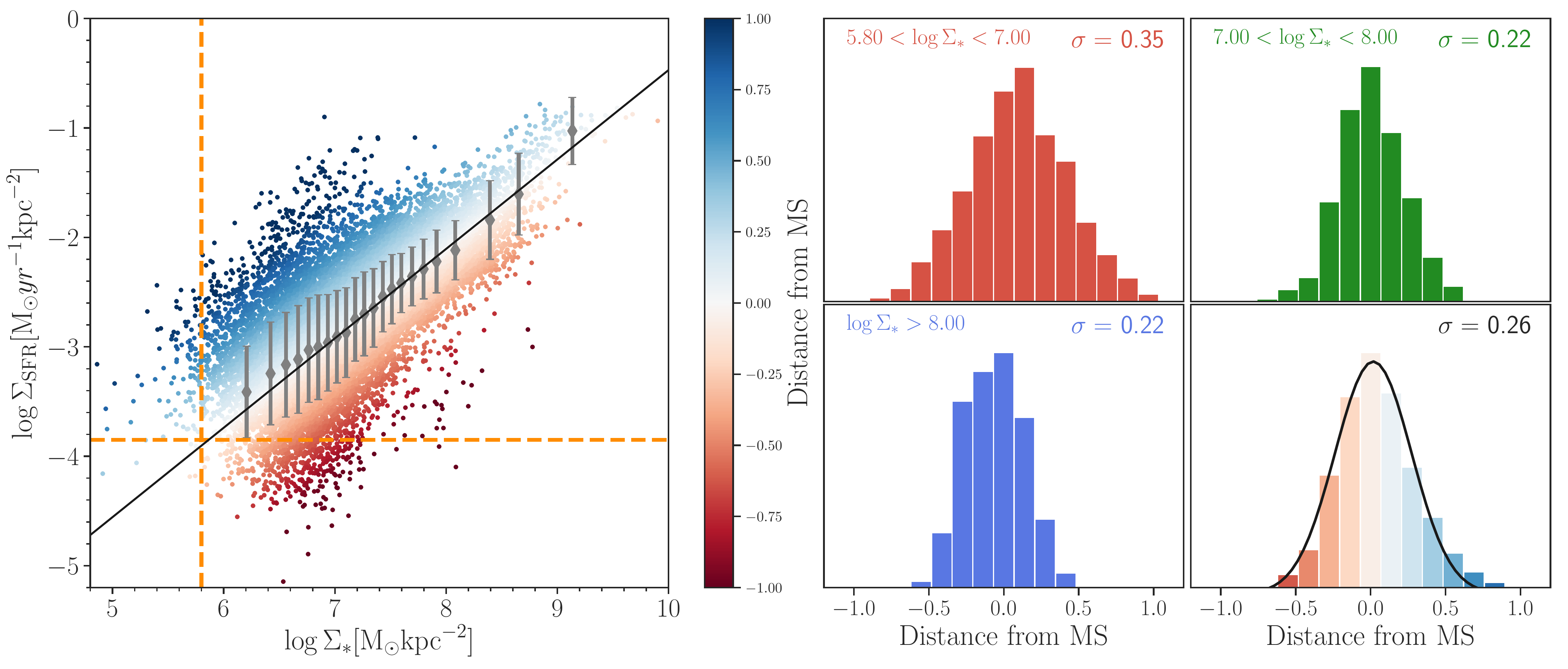}
    \caption{The fitted main sequence, and its scatter. {\it Left panel:}  8$\arcsec$ cells results for star formation rate surface density and stellar mass density, color coded as a function of their distance from the main sequence. Orange dashes lines are the sensitivity limits. The MS (black solid line) is obtained as the linear fit of the gray data points. The error bars are the 1$\sigma$ dispersion in each bin. {\it Right panel:} distribution of distances from the MS, in three stellar mass bins (red, cyan and blue, color coded as in the left panel). The $\sigma$ values reported in each panel are obtained by fitting the distribution with a Gaussian. The scatter varies between 0.22 for the highest mass bin to 0.35 for the lowest mass bin.}
    \label{fig:MS_scatter}
\end{figure*}
In Fig.\,\ref{fig:GlobalProp} we also show the MS relation of \citet[][dashed purple line]{2014ApJ...795..104W}, that was originally computed using integrated galaxy values in the redshift range [0.5:1.1] and for sources with stellar masses between 8.5 logM$_{\odot}$ and 11.5 logM$_{\odot}$. We also show the scatter of the MS relation derived for our sample (shaded region) at z $\sim 0$, as more evidence is mounting up for a non evolution of the MS scatter with redshift \citep[e.g.][]{2019MNRAS.490.5285P}. We rescale the Whitaker MS using the redshift evolution given in \citet{2014ApJ...795..104W}. Interestingly, despite the different procedures and different stellar mass scales probed in the two works, the MS of \citet{2014ApJ...795..104W} is extremely consistent with the one found in this paper. This agreement clearly indicates the existence of the integrated MS relation as a consequence of a process that regulates the SF activity on small-scales.

We then computed the MS relation in the log$\Sigma_{\star}$ - log$\Sigma_{\rm SFR}$ plane considering all the accepted cells coming from the 8 galaxies. The fit, a log-linear relation $\log \Sigma_{\rm SFR} = m \log \Sigma_{\star} + q$, is performed using {\sc emcee} \citep{2013PASP..125..306F}. Since we are dealing with a large number of points (from $\sim 3000$ for 1.5 kpc apertures to $\sim 30000$ for 8$\arcsec$ cells), we bin the data and perform linear fit on the binned data points (gray points in Fig.\,\ref{fig:MS_scatter}). Whenever binning, we discard the points falling below the sensitivity limits and consider only the ones above. The mass bins are built in two steps: first we subdivide the $\log \Sigma_{\star}$ interval in 20 bins, each containing the same number of points. In order to probe the upper part of the MS (that otherwise would not have any binned points) we add a bin between $8.5 \leq \Sigma_{\star} < 9$ and one over 9 M$_{\odot}$ yr$^{-1}$ kpc$^{-2}$. Within each bin, we compute the median $\Sigma_{\star}$ and $\Sigma_{\rm SFR}$, while the error is taken as the standard deviation of the points inside the bin. We fit the binned data points with a $\log$-linear relation. In Fig.\,\ref{fig:MS_scatter} we report the 8$\arcsec$ cells results for the sample, and the fit to the data (left panel). The results of the fit are a slope of $0.82 \pm 0.12$ and intercept of $-8.69  \pm 0.97$. As expected from Fig.\,\ref{fig:SFR_comparison}, we tested that the MS obtained fitting the SFRs coming directly from {\sc MAGPHYS} is consistent (almost identical) with this one.

Our slope is consistent with others in literature (see Sec.\,\ref{sec:comparison}) that usually span the range $0.7-1.1$ despite using different SFR tracers and different redshift and stellar mass ranges for the analysis \citep[i.e.][]{2013ApJ...764L...1P, 2013ApJ...779..135W, 2014ApJ...797..108H, 2016ApJ...821L..26C, 2016A&A...590A..44G, 2017MNRAS.469.2806A, 2018MNRAS.479.5083A, 2018ApJ...865..154H, 2019MNRAS.488.3929C}. It is worth noting that several of the literature works implement the orthogonal distance regression (ODR) method to find the location of the MS relation. If we perform an ODR fit on our results, we obtain a slope of 0.89 and an intercept of -9.13.

The right panels of Fig.\,\ref{fig:MS_scatter} are dedicated to the distribution scatter $\sigma$. We investigate if the scatter of the spatially resolved MS varies as a function of mass by dividing our sample in three mass bins, $5.8 \leq \log \Sigma_{\star} [{\rm M}_{\odot} {\rm yr}^{-1}] < 7.0$ (red histogram), $7.0 \leq \log \Sigma_{\star} [{\rm M}_{\odot} {\rm yr}^{-1}]  < 8.0$ (green histogram), $\log \Sigma_{\star} > 8.0 {\rm M}_{\odot}$ (blue histogram). The total scatter of the full sample (colored histogram) is $\sigma = 0.26$. There are hints of a decreasing scatter with increasing stellar mass, from 0.35 to 0.23. Once again, these values fall within the typical ranges of 0.15-0.35 reported in the literature for the spatially resolved MS \citep{2016ApJ...830...83C, 2016MNRAS.456.4533M, 2017MNRAS.466.1192M, 2018ApJ...865..154H}. The importance of the scatter and the way it relates with the gas properties is explored in Morselli et al. (in prep).

\begin{figure}
	\includegraphics[width=\columnwidth]{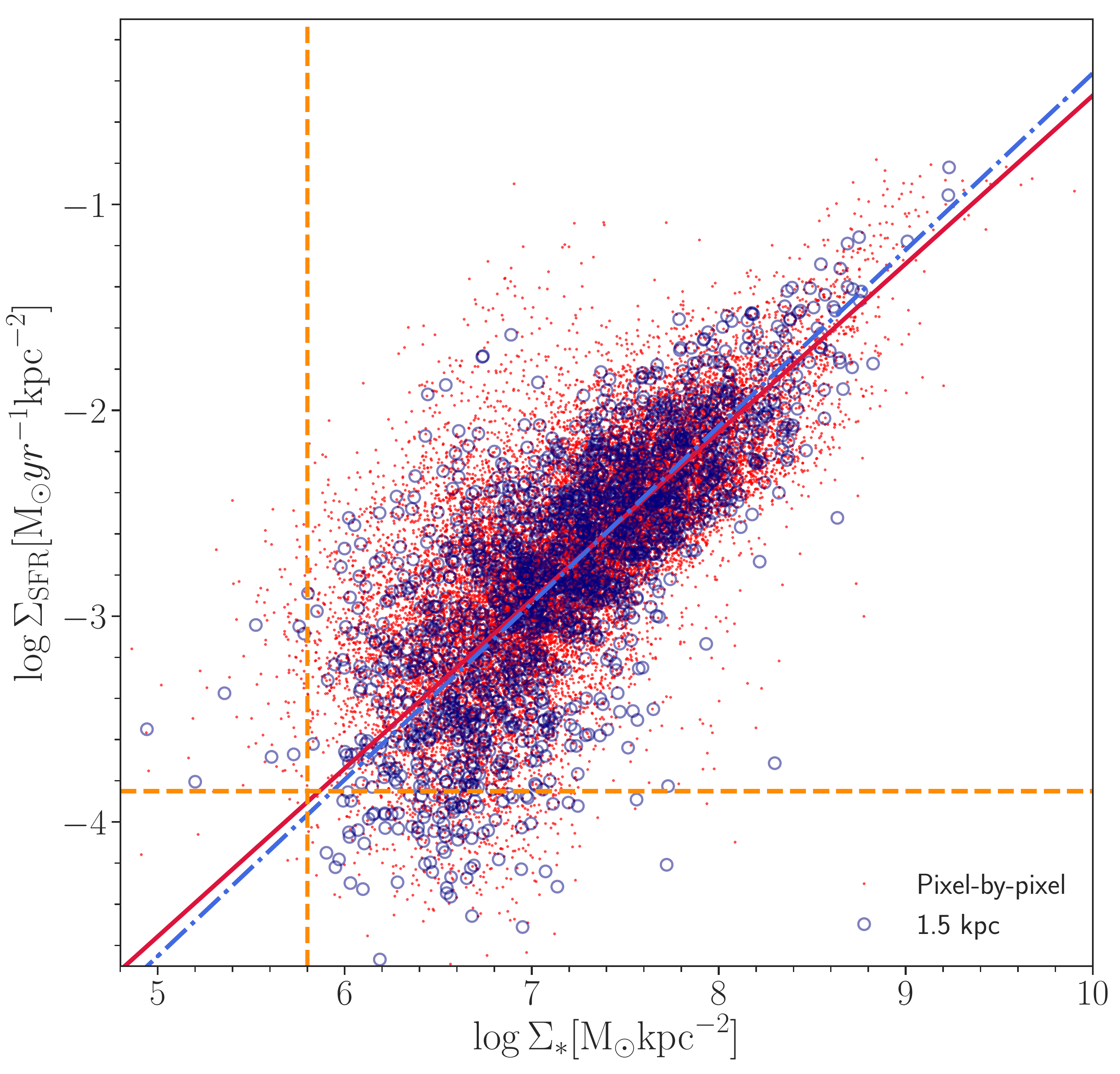}
    \caption{$\Sigma_{*}$-$\Sigma_{\rm SFR}$ results for 8$\arcsec$ (red points) and 1.5 kpc (blue circles) cells. Dashed lines are the sensitivity thresholds. It is clear how the results coming from the higher 1.5 kpc physical scale are the mean of the properties coming from the single cells inside them. Both data sets lead to the same linear relations (red line and blue dashed line respectively), well within the uncertainties.}
    \label{fig:comparison_plot}
\end{figure}

Finally, we check that the results are consistent on different spatial scales. In Fig.\,\ref{fig:comparison_plot} we compare the surface densities obtained for 8$\arcsec$ cells (scales varying from 0.2 to 0.8 kpc) and 1.5 kpc cells. The black solid line is the MS fit to 8$\arcsec$ data (though the fit to 1.5 kpc data lead to the same values well within the uncertainties), orange lines the sensitivity thresholds. Both distributions follow the same relation, and occupy the same region of the plane, being clear how the retrieved surface density properties on bigger scales reflects the average surface density properties in enclosed smaller scales. The main differences arise for galaxy regions where $\chi^2$ threshold is not reached, as expected since 1.5 kpc have an intrinsically higher signal-to-noise ratio for the SED points, always in regions below the sensitivity thresholds. The fact that the two realizations in different physical scales (sometimes even of an order of magnitude, i.e. NGC5457) are almost overlapping tells us that the star formation laws still hold on physical scales lower than $\sim 500$ pc.

\subsection{Comparison with other resolved star-forming main sequences}\label{sec:comparison}
In recent years, as already mentioned in the introduction, several works have analysed the spatially resolved MS of SFGs, thanks to the increased availability of integral field spectroscopic data. Surveys like MaNGA, CALIFA and SAMI made it possible to obtain the SFR from the H$\alpha$ luminosity, and to correct it for dust absorption through the Balmer decrement. The lack of information on the IR emission, however, might prevent a comprehensive evaluation of the energetic budget in galaxies, as UV and optical tracers could underestimate a fraction of the obscured SFR \citep[e.g.][]{2014MNRAS.443...19R}. In this Section, we thus compare our panchromatic results with the spatially resolved MS relations obtained with data from CALIFA \citep{2016ApJ...821L..26C}, MaNGA \citep{2017ApJ...851L..24H, 2019MNRAS.488.3929C, 2019MNRAS.tmp.2839B}, and SAMI \citep{2018MNRAS.475.5194M}. For completeness, we also compare our results with the MS relations of \citet{2017MNRAS.469.2806A}, that perform pixel-by-pixel SED fitting to GALEX and SDSS photometry of local ($0.01 < z < 0.02$) massive spiral galaxies selected in the MPI-JHU (Max Planck Institute for Astrophysics-Johns Hopkins University), and of \citet{2018ApJ...865..154H}, obtained for a sample of 355 nearby galaxies, with spatially resolved observations of H$\alpha$ and mid-IR emission. 

\begin{figure}
	\includegraphics[width=\columnwidth]{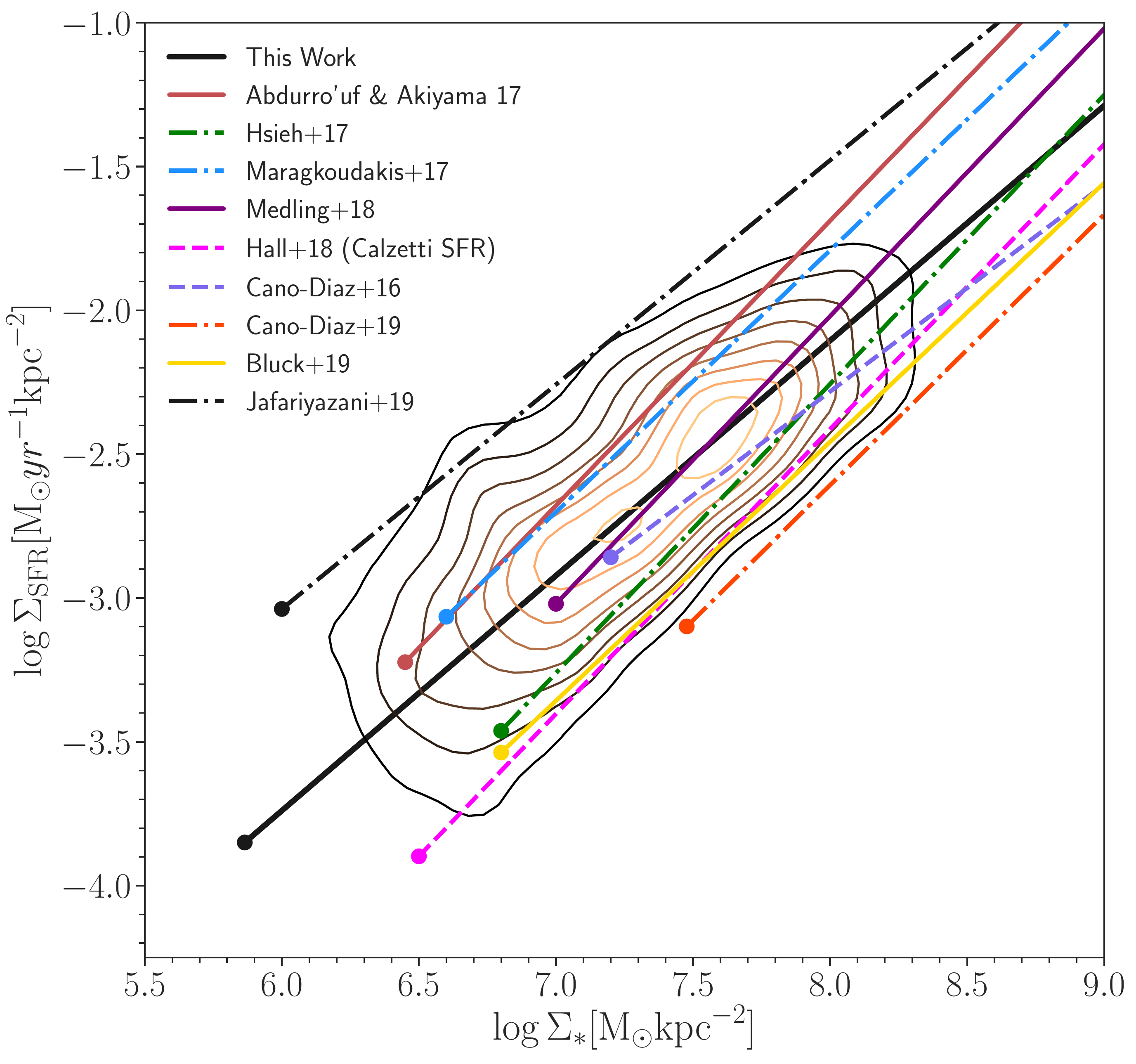}
    \caption{A comparison between our results (black-to-yellow contours encircling from 80\% to 10\% of our data points, fitted by the solid black line) and a sample of resolved MS relations from the literature, respectively. Each relation, that has been converted to a \citet{2003PASP..115..763C} IMF, and re-scaled to z = 0 (as described in the text), starts from the sensitivity limit reported in the study (or, if not, estimated from the relative data).}
    \label{fig:MS_relations}
\end{figure}

In Fig.\,\ref{fig:MS_relations} we show the spatially resolved MS relation of this work (solid black line) together with the relations mentioned above. To underline the depth of each study, in Fig.\,\ref{fig:MS_relations} we plot the relations as starting from the logM$_{\star}$ value above which 80$\%$ of the corresponding data are located. The relations have been homogenised in terms of IMF and cosmology, and rescaled to the median redshift of our sample, using the evolution in MS normalisation of SFR $\propto$ (1+z)$^{2.88}$ \citep{2014ApJ...795..104W}.

\begin{table}
    \centering
    \caption{Summary of the slopes quoted in the text, with associated SFR tracer.}
    \label{tab:MS_comparison}
    \begin{tabular}{lcc}
        \hline
        Reference                & Slope & SFR tracer \\
        \hline
        \citet{2016ApJ...821L..26C} & 0.72  & H$\alpha$ \\
        \citet{2017MNRAS.469.2806A} & 0.99  & SED fitting \\
        \citet{2017ApJ...851L..24H} & 1.00  & H$\alpha$ \\
        \citet{2017MNRAS.466.1192M} & 0.91  & 3.6 $\mu$m or 8.0 $\mu$m   \\
        \citet{2018ApJ...865..154H} & 0.99  & H$\alpha$ + 24$\mu$m   \\
        \citet{2018MNRAS.475.5194M} & 1.00  & H$\alpha$ \\
        \citet{2019MNRAS.488.3929C} & 0.94  & H$\alpha$ \\
        \citet{2019MNRAS.tmp.2839B} & 0.90 & H$\alpha$ \\
        This Work & 0.82 & SED fitting \\
        \hline
    \end{tabular}
\end{table}
The spatially resolved MS of \citet[][lavender dashed line]{2016ApJ...821L..26C} has been estimated from a sample of 306 galaxies with mixed morphologies at $0.005 < z < 0.03$, on spatial scales of 0.5-1.5 kpc. They find a slope of 0.72. Similarly, \citet{2017ApJ...851L..24H} used 536 SFGs at $0.01 < z < 0.15$ from the MaNGA survey to obtain a MS relation on scales of $\sim$1 kpc (green dash-dotted line), obtaining a slope of 1.00 with the ODR fitting method. Recently, \citet{2019MNRAS.488.3929C} updated the work of \citet{2017ApJ...851L..24H} using $\sim$ 2000 galaxies from the MaNGA MPL-5 release ($0.01 < z < 0.15$) to probe the spatially resolved MS for a larger sample, again on spatial scales of $\sim$ 1.0 kpc. Their MS relation (orange dash-dotted line) has a slope of 0.94. Based on $\sim 3500$ local galaxies in the SDSS-IV MaNGA-DR15 data, with SFRs coming from D4000 and H$\alpha$ observations whenever available, \citet[][yellow line]{2019MNRAS.tmp.2839B} obtain a slope of 0.90 on a dataset of over 5 million spaxels. \citet[][blue dotted line]{2017MNRAS.466.1192M} evaluate SFRs from IRAC 3.6 and 8.0 $\mu$m data at sub-kpc scales on a sample of 369 nearby galaxies ($z \sim 0.02$), obtaining a slope of 0.91. The MS of \citet[][purple solid line]{2018MNRAS.475.5194M} has been estimated from $\sim$ 800 galaxies in the SAMI Galaxy Survey at $z < 0.1$ and has a slope of 1.0. The MS relation of \citet[][magenta dashed line]{2018ApJ...865..154H} has a slope of 0.99, and was computed exploiting spatially resolved H$\alpha$ observations of nearby galaxies within the Survey of Ionization in Neutral Gas Galaxies (SINGG) and WISE surveys
\citep[we report here the relation obtained using the SFR transformation of][]{2007ApJ...666..870C}. Finally, the MS of \citet[][red line]{2017MNRAS.469.2806A} has a slope of 0.99.

The immediate take-home information evident from Fig.\,\ref{fig:MS_relations} is that thanks to our multiwavelength approach we are able to probe regions of lower stellar mass surface densities (up to a factor of 10) compared to spectroscopic observations, bound to the sensitivity limit of the H$\alpha$ line. This translates in the ability of sampling regions that are located further away from the galaxy centre, up to $\sim$ 2.5 - 3 effective radii (Re). While it is true that CALIFA galaxies have been selected in size so that most of them are fully sampled by the instrument field of view, the requirement to observe the H$\alpha$ line with a signal-to-noise ratio (S/N) > 3 in each resolution element translates in smaller radii within which the SFR can be efficiently computed, as it is clear from the different sensitivity limit. The same reasoning applies to the MS relations of \citet{2018MNRAS.475.5194M} and \citet{2019MNRAS.488.3929C}. On the other hand, \citet{2018ApJ...865..154H} reach lower $\Sigma_{\rm SFR}$ thanks to the combination of H$_{\alpha}$ and mid-IR data, but has the drawback that low $\Sigma_{\rm SFR}$ only trace obscured SF. 

The MS slope that we obtain is lower than the ones shown in Fig.\,\ref{fig:MS_relations} with the exception of the \citet{2016ApJ...821L..26C} relation. To exclude effects due to different fitting procedures, that can largely affect the MS slope and intercept, we also compute the relation using the ODR method, and we find a slope of 0.88. This value is still lower than the ones found in the works mentioned above, but it is important to underline that our relation was computed from 8 galaxies against the hundreds of sources of other works. As there are strong galaxy-to-galaxy variations, it is important to apply our approach on a larger sample to carry out a more meaningful comparison with other works. 

\section{Discussion and conclusions}
In this paper we have presented a new analysis of eight local face-on spirals, with the aim of understanding if the MS is a universal relation holding also on sub-galactic scales. Other surveys of nearby galaxies have addressed this question, showing clear correlation between stellar masses and star-formation per unit area, and to their gas content \citep{2008AJ....136.2846B,  2008AJ....136.2782L, 2015A&A...577A.135C}. By exploiting the publicly available photometric information of DustPedia in the UV to far-Infrared spectral range, we have been able to perform a global SED fitting procedure that allowed us to simultaneously account for a careful evaluation of the obscured and unobscured SFR components, over the full optical radius, larger than what obtained with optical IFS at similar redshifts. Our limited sample is restricted to the grand design spirals with low inclination, large spatial extension and regular spiral arms structures in DustPedia. This analysis has provided a total of tens of thousands physical cells on typical scales of $\sim0.5$ kpc over very different internal galaxy environments (bulges, spiral arms, inter-arms regions, outskirts). This set is thus well fit to study the secular evolution processes that regulate the star formation in local sources, dominated by rotationally supported systems \citep[e.g.][]{2009ApJ...697.2057L, 2009ApJ...706.1364F, 2013PASA...30...56G, 2015ApJ...799..209W, 2017ApJ...843...46S, 2018ApJS..238...21F, 2019ApJ...880...48U}.

Some individual galaxies show peculiar variation around the $\Sigma_{\rm SFR}-\Sigma_{\star}$ relation presented in Fig.\,\ref{fig:MS_scatter} (see bottom-right panels in Figs A1-A8). For example, NGC3938 and NGC4254 show an average enhancement of the SFR, that we interpret as a possible effect of the environment, as these two sources are located in the Ursa Major group and the Virgo cluster, respectively. Other galaxies, in particular NGC0628, show a prominent bulge feature appearing as a narrow distribution at the highest stellar mass densities. Indeed, we already mentioned that at the lowest stellar mass densities (i.e. at large galactocentric distances) in few sources we have identified a cloud of cells deviating from the main relation toward lower SFR, at fixed stellar mass (NGC4321 is the cleanest example). Such distributions are circularly distributed in the outer parts of the galaxies, out of the regions spanned by the dynamical interaction of the  spiral arms. It is still to be understood if this is associated to an older population migrated out of the disk, or if it is the remnant of external accretion through, for example, minor merging. Finally, only NGC5457 reveals strong mini-starbursts inside the spiral arms (i.e. regions well elevated above the MS, by a factor larger then 10-100 times at fixed mass density).

In any case, the combination of all the eight galaxies demonstrates the existence of a universal relation, as the deviation of single sources is well within the global scatter of $\sim 0.27$ dex (see Sect. 4.3). We have then demonstrated that such relation holds at different galaxy scales, supporting the interpretation from other surveys that the SFR is regulated by local processes of gas-to-stars conversion happening at GMC scales on the order of a few hundred pc. With respect to other works, we have however provided evidence that such {\em secular regulation} keeps to apply at the farthest galactocentric distances, where the optical disk is still influenced by the density waves motions, originating  the spiral arms.

The present work defines an accurate locus in the $\Sigma_{\rm SFR}-\Sigma_{\star}$ plane, constrained by observed distributions covering 3 dex in the parameters space. Such a definition of the spatially resolved MS in local disk galaxies represents a valuable reference for future comparison to different galaxy morphological types adopting a similar panchromatic approach. Indeed, there are indication that galaxy morphology plays an important role in the characterising the spatially resolved MS relation, especially its scatter \citep[e.g.][]{2016A&A...590A..44G, 2019MNRAS.488.3929C}. \citet{2017MNRAS.466.1192M} observe a decrease in the spatially resolved MS relation from late-type to early-type spirals, while the scatter remains constant. Other works do not find a similar connection \citep{2018ApJ...865..154H}. These examples show the importance of extending the analysis performed in this work to a larger sample of galaxies encompassing different morphologies.

In the second paper of this series, we will exploit this reference sample to study if and how the distance to the resolved MS is connected to the total (atomic and molecular) gas content (Morselli et al., in prep.).

\section{Summary}

The star-forming main sequence is a well studied tight relation between stellar masses and star formation rates, observed up to $z \sim 6$, over a great variety of environments and morphologies, both globally, counting galaxies as a whole, and locally, resolving physical properties in single galaxy regions. In this work we perform spatially resolved SED fitting in a sample of 8 local grand-design spirals taken from the DustPedia archive and we use the outputted maps of stellar mass and star formation rate (see Appendix) to analyse the spatially resolved MS of SFGs on scales spanning the range between 0.4-1.5 kpc. We summarise here our main findings: 

\begin{itemize}
\item When considering the 8 galaxies together, we obtain a spatially resolved Main Sequence with a slope of 0.82 and an intercept of -8.69. When fitting the data with the ODR method we obtain a slope and an intercept of 0.88 and -9.05, respectively. This relation holds on different scales, from sub-galactic to galactic;

\item The local spatially resolved MS is consistent with the evolutionary (from high-z) low-mass relation, thus proving its universality across cosmic time. This is a crucial point to validate the integrated information on individual galaxies at high redshift;

\item We have overtaken the limits of all the spectroscopic resolved emission lines studies, based mainly on H$\alpha$ and Balmer decrement corrections for dust extinction. The sensitivities of these surveys do not sample the lowest M$_{\star}$ and SFR as we do with a multiwavelength photometric approach, allowing us to probe the outermost regions of galaxies.
\end{itemize}

We plan to extend this analysis on different morphological types inside the DustPedia sample, while improving the SED fitting pipeline to investigate star formation histories of resolved galaxy regions. We will also link these results with existing gas observations (i.e. CO, H$_2$, CII) to further understand the role of secular evolution in galaxies.

\section*{Acknowledgements}
We would like to thank the anonymous referee for the useful comments improving the overall work quality. We are grateful to CJR Clark for the information on CAAPR, and to M. Cano-Diaz for providing us the CALIFA results from her 2016 paper. AE and GR are supported from the STARS@UniPD grant. GR acknowledges the support from grant PRIN MIUR 2017 - 20173ML3WW\_001. GR and CM acknowledge funding from the INAF PRIN-SKA 2017 program 1.05.01.88.04. We acknowledge funding from the INAF main stream 2018 program "Gas-DustPedia: A definitive view of the ISM in the Local Universe". This research is based on observations made with the Galaxy Evolution Explorer, obtained from the MAST data archive at the Space Telescope Science Institute, which is operated by the Association of Universities for Research in Astronomy, Inc., under NASA contract NASA 5-26555. DustPedia is a collaborative focused research project supported by the European Union under the Seventh Framework Programme (2007-2013) call (proposal no. 606847). The participating institutions are: Cardiff University, UK; National Observatory of Athens, Greece; Ghent University, Belgium; Universit\'e Paris Sud, France; National Institute for Astrophysics, Italy and CEA (Paris), France. This research made use of Photutils, an Astropy package for detection and photometry of astronomical sources.

\bibliographystyle{mnras}
\bibliography{biblio}

\appendix

\section{Galaxy-by-galaxy results}\label{App:results}
In this Appendix, we present the individual galaxy-by-galaxy results, reporting the SFR and stellar mass maps, along with a map measuring the distance from the fitted main sequence (blue points towards starburst, red points towards quenched ones). In the lower left panel, we report the results on the $M_{\star}$-SFR plane, with fitted (total) MS in black and the one fitting the lone galaxy results.

\begin{figure*}
	\includegraphics[width=\columnwidth]{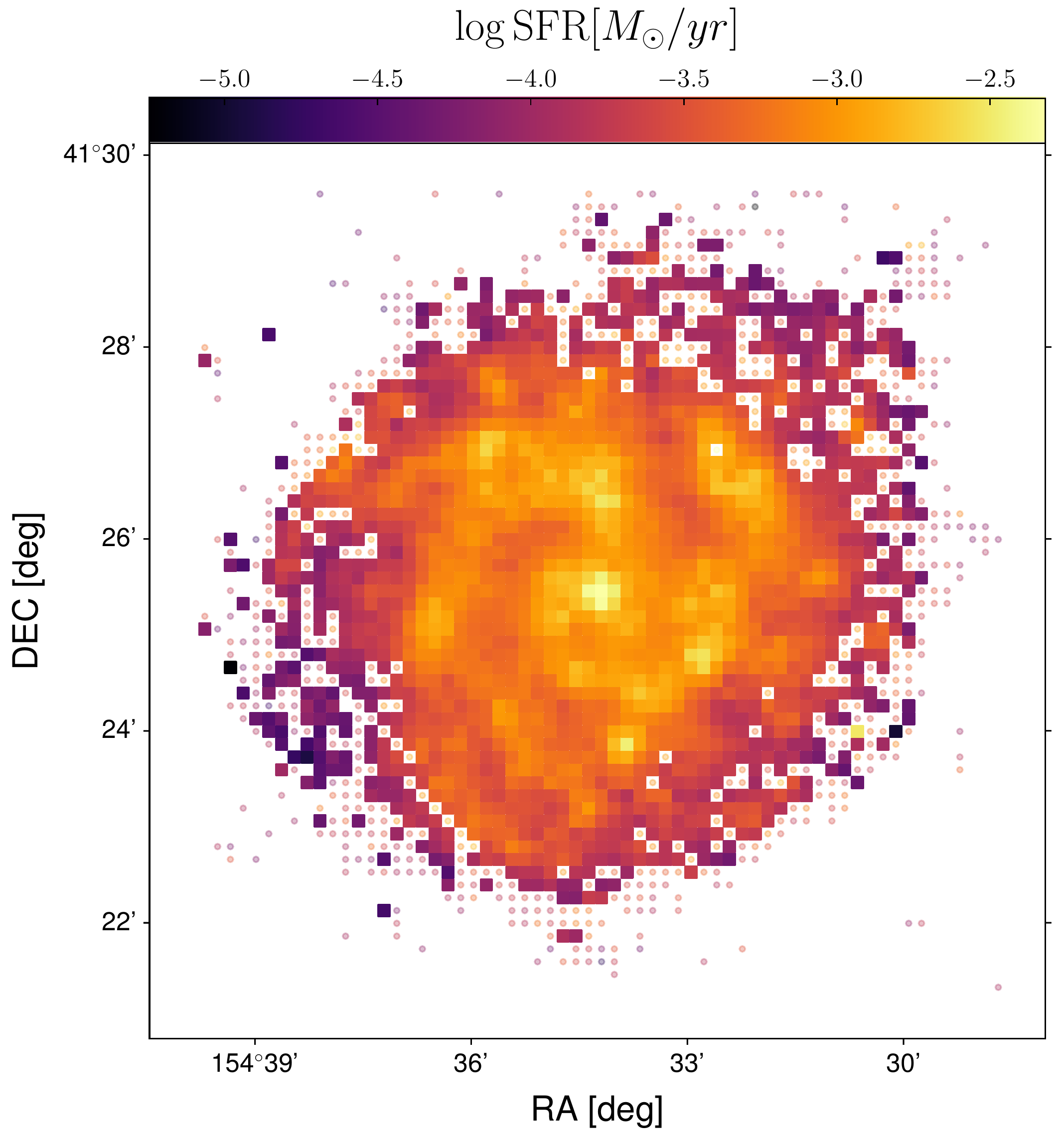}
	\includegraphics[width=\columnwidth]{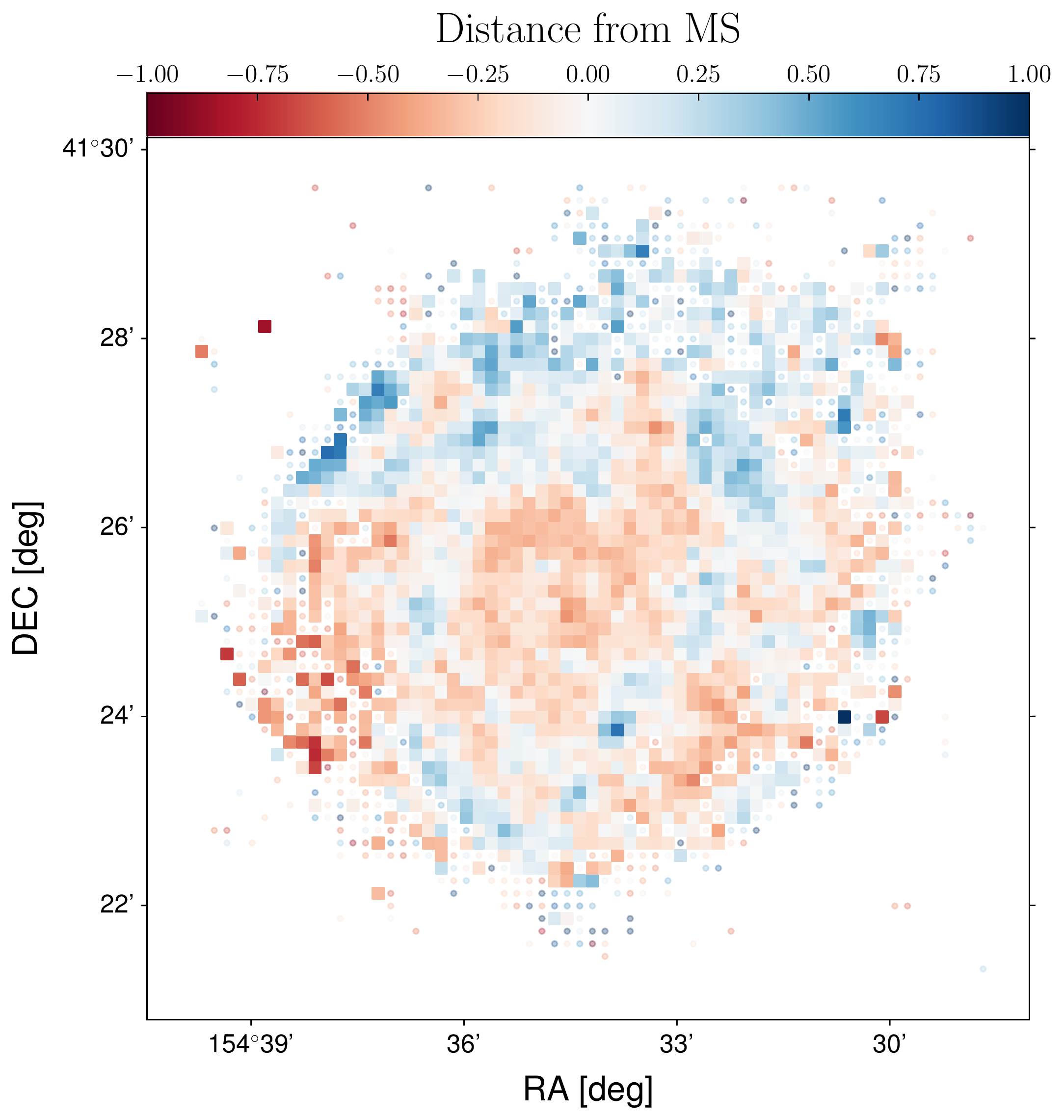}
	\includegraphics[width=\columnwidth]{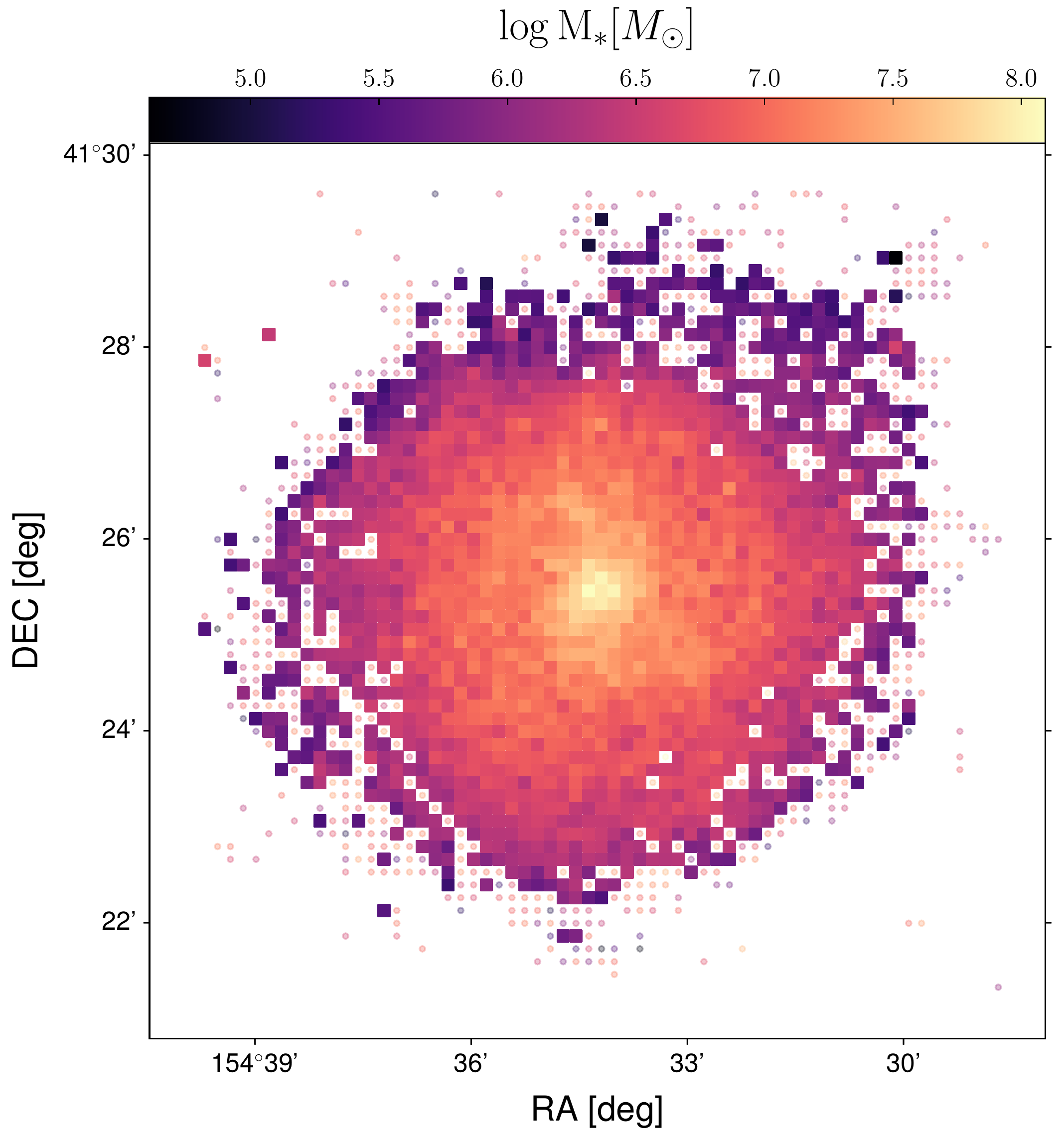}
	\includegraphics[width=1.02\columnwidth]{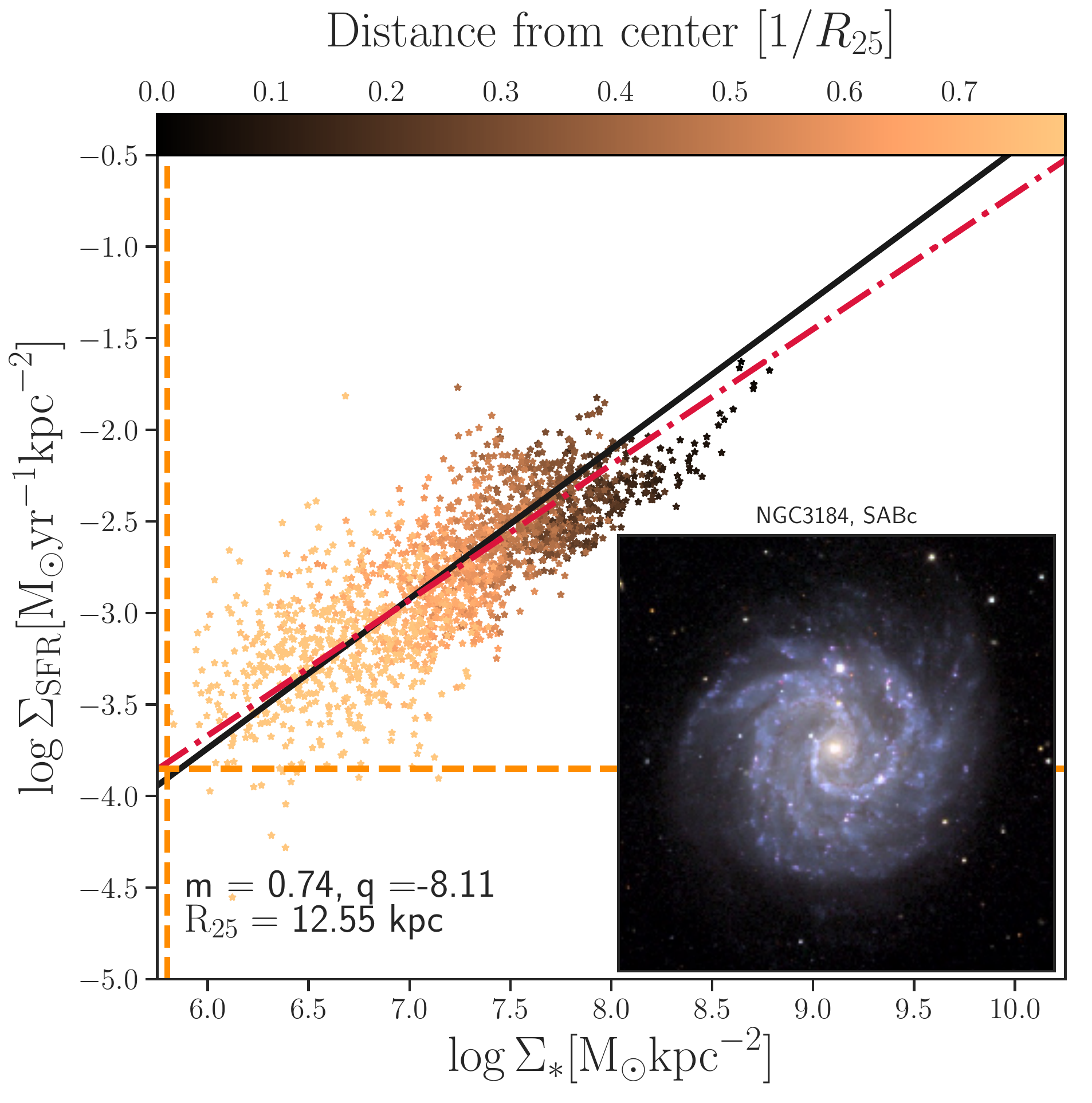}
    \caption{Same as Fig.\,\ref{fig:NGC0628}, for NGC3184}
    \label{fig:NGC3184}
\end{figure*}

\begin{figure*}
	\includegraphics[width=\columnwidth]{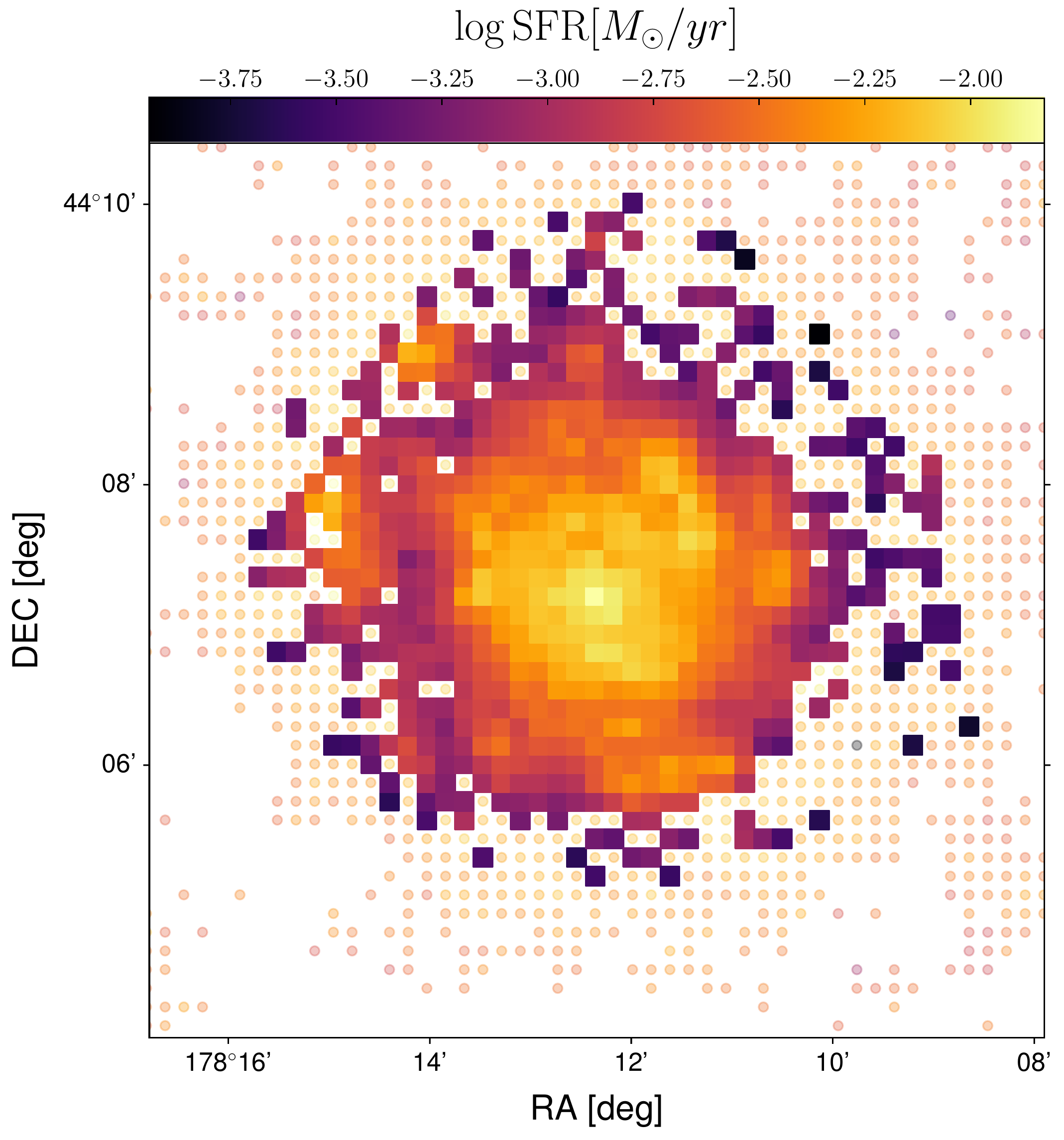}
	\includegraphics[width=\columnwidth]{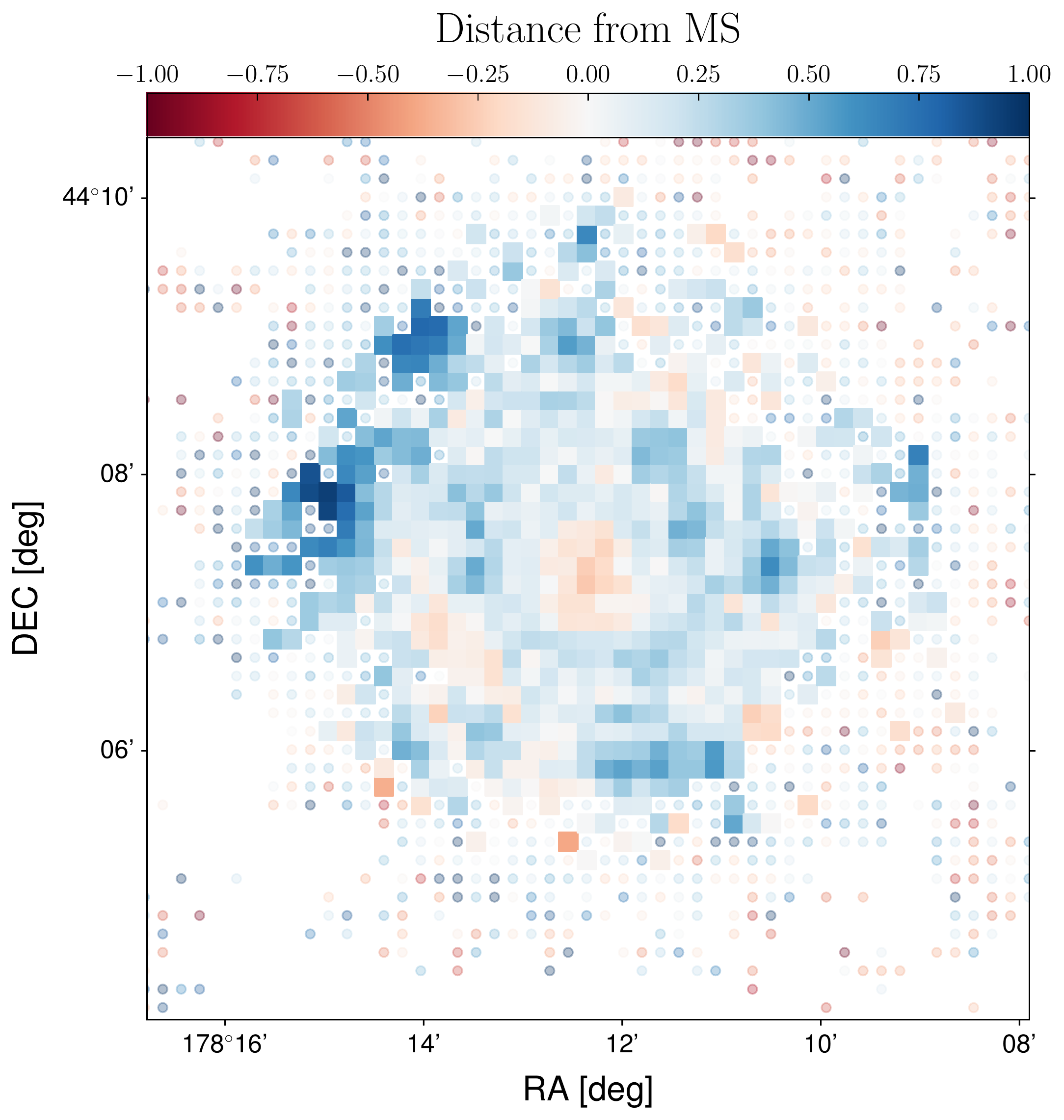}
	\includegraphics[width=\columnwidth]{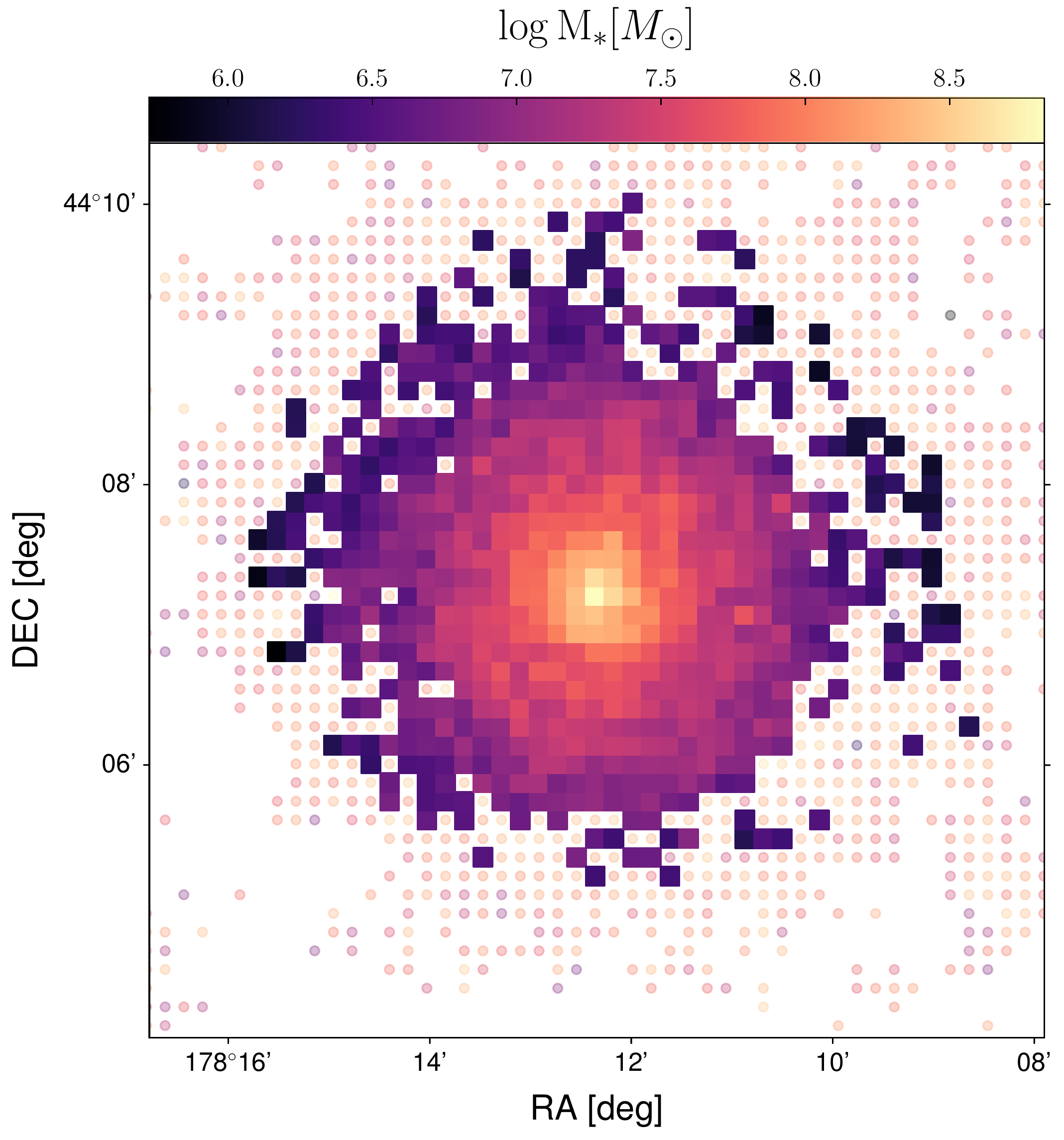}
	\includegraphics[width=1.02\columnwidth]{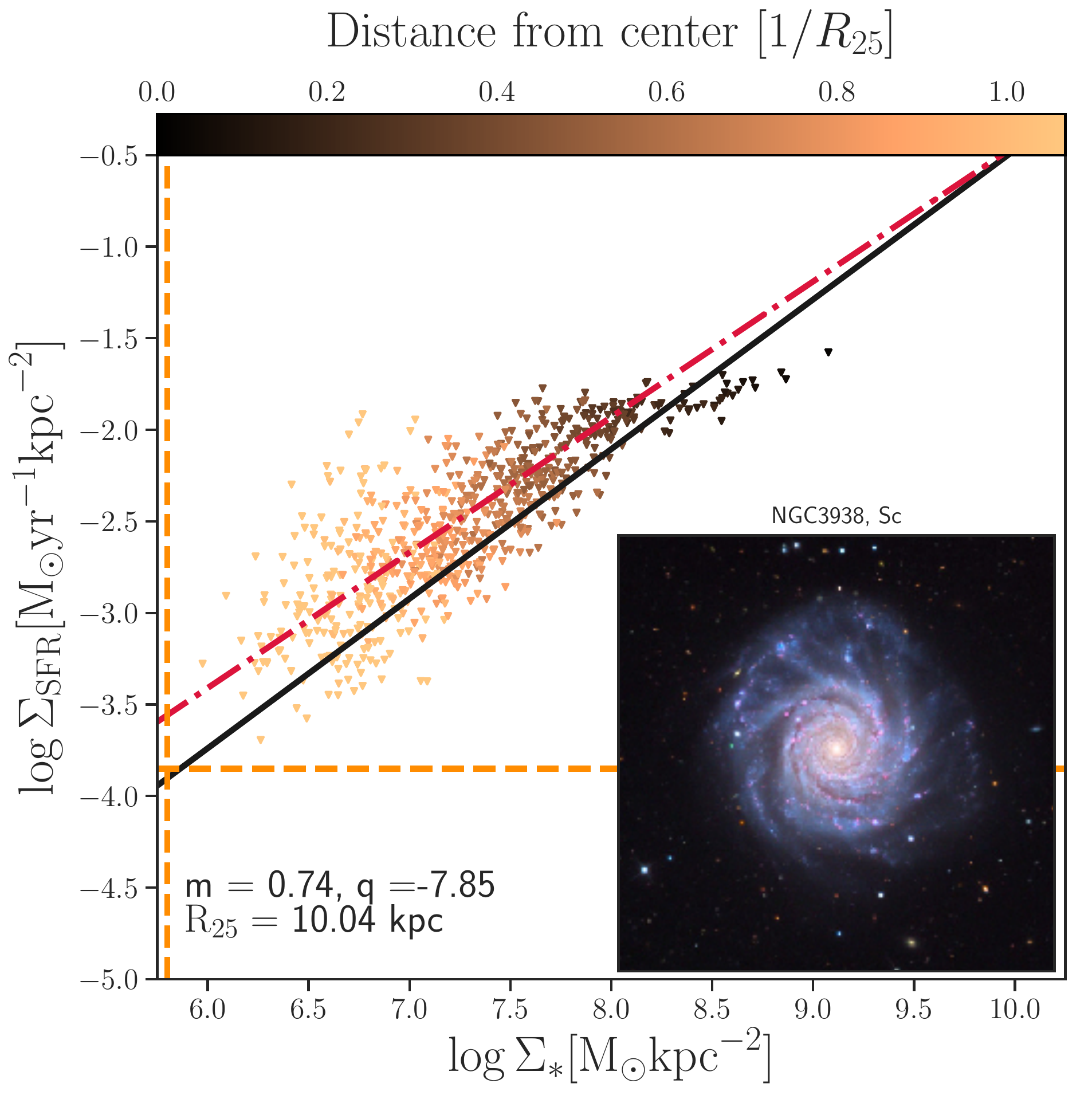}
    \caption{Same as Fig.\,\ref{fig:NGC0628}, for NGC3938.}
    \label{fig:NGC3938}
\end{figure*}

\begin{figure*}
	\includegraphics[width=\columnwidth]{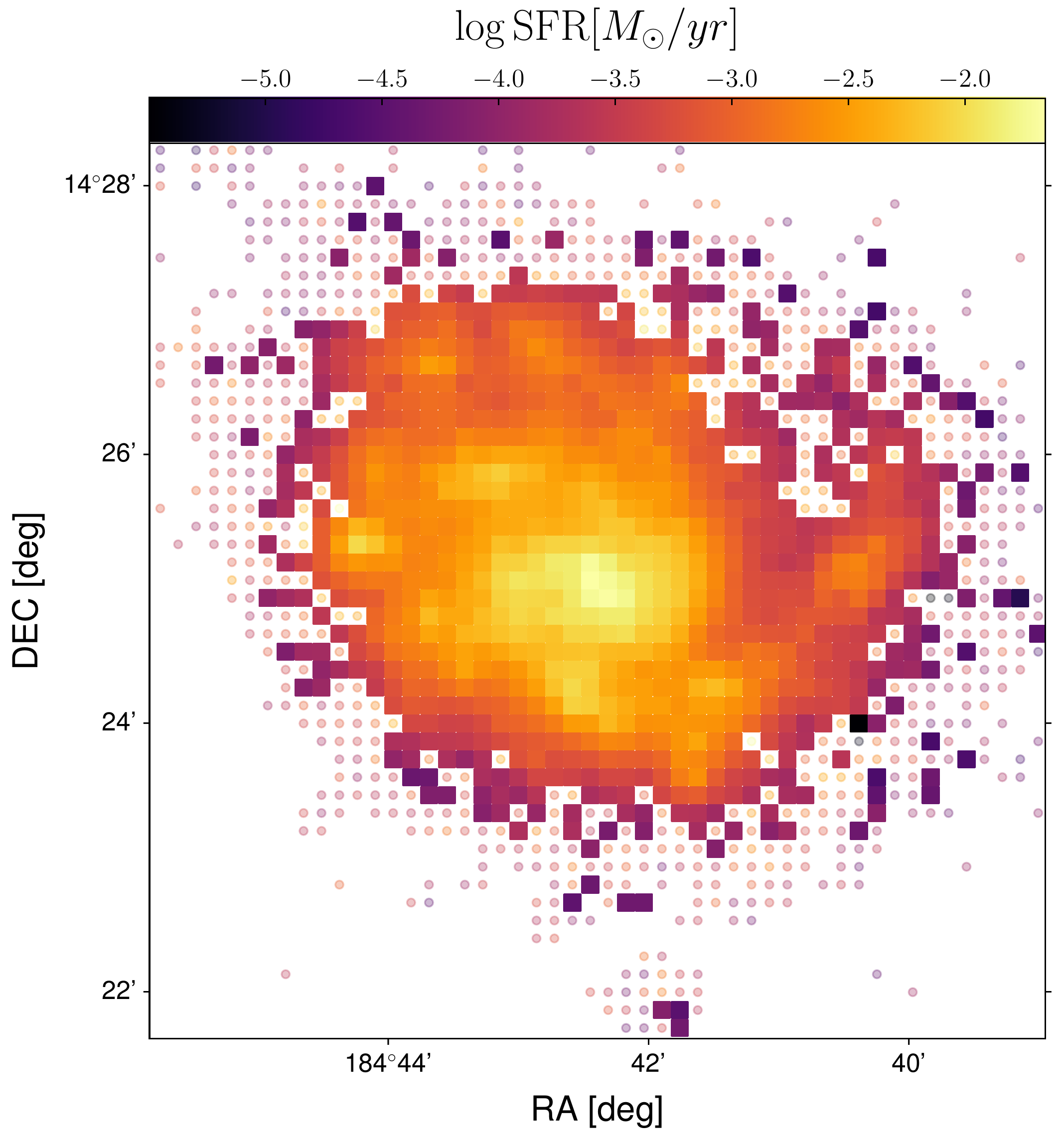}
	\includegraphics[width=\columnwidth]{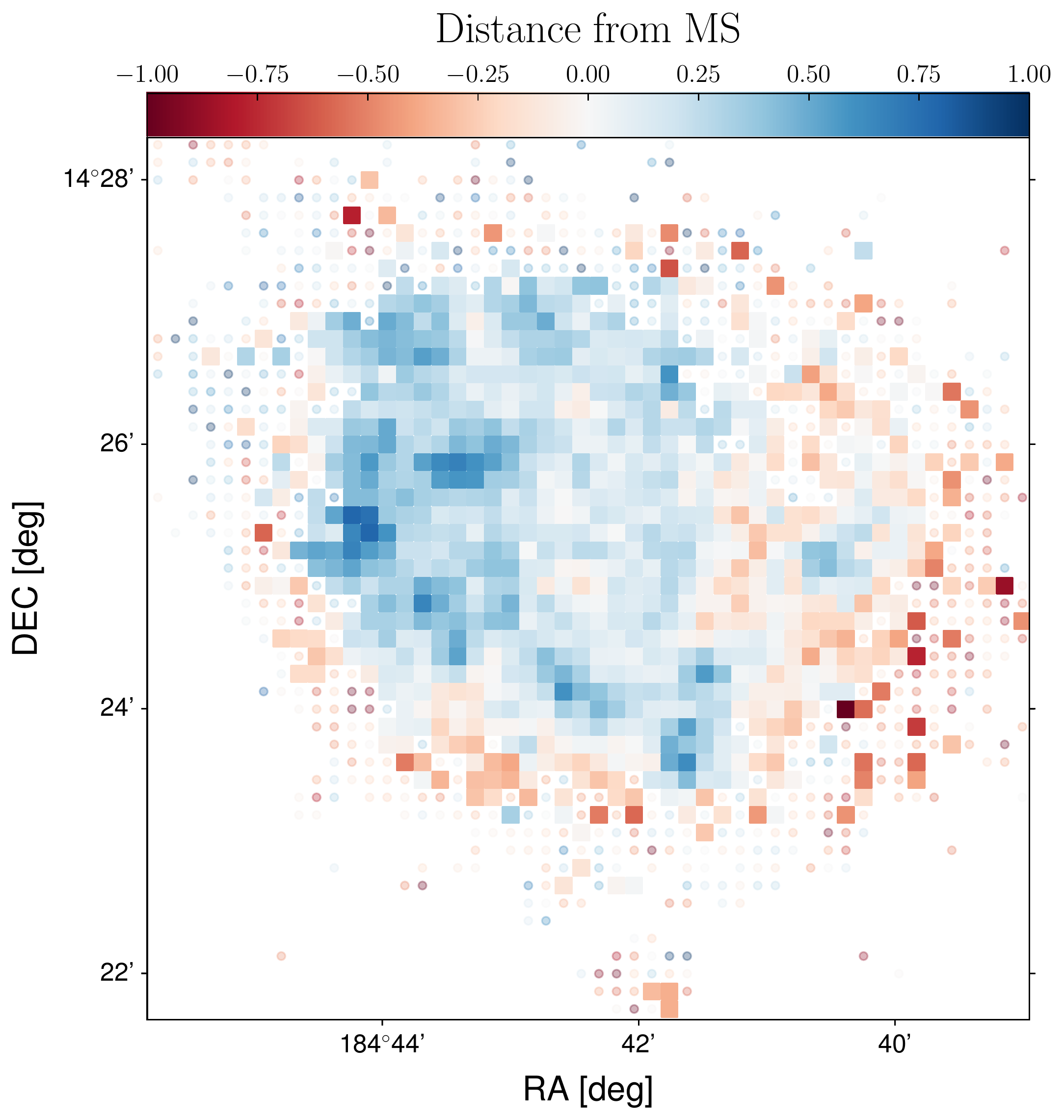}
	\includegraphics[width=\columnwidth]{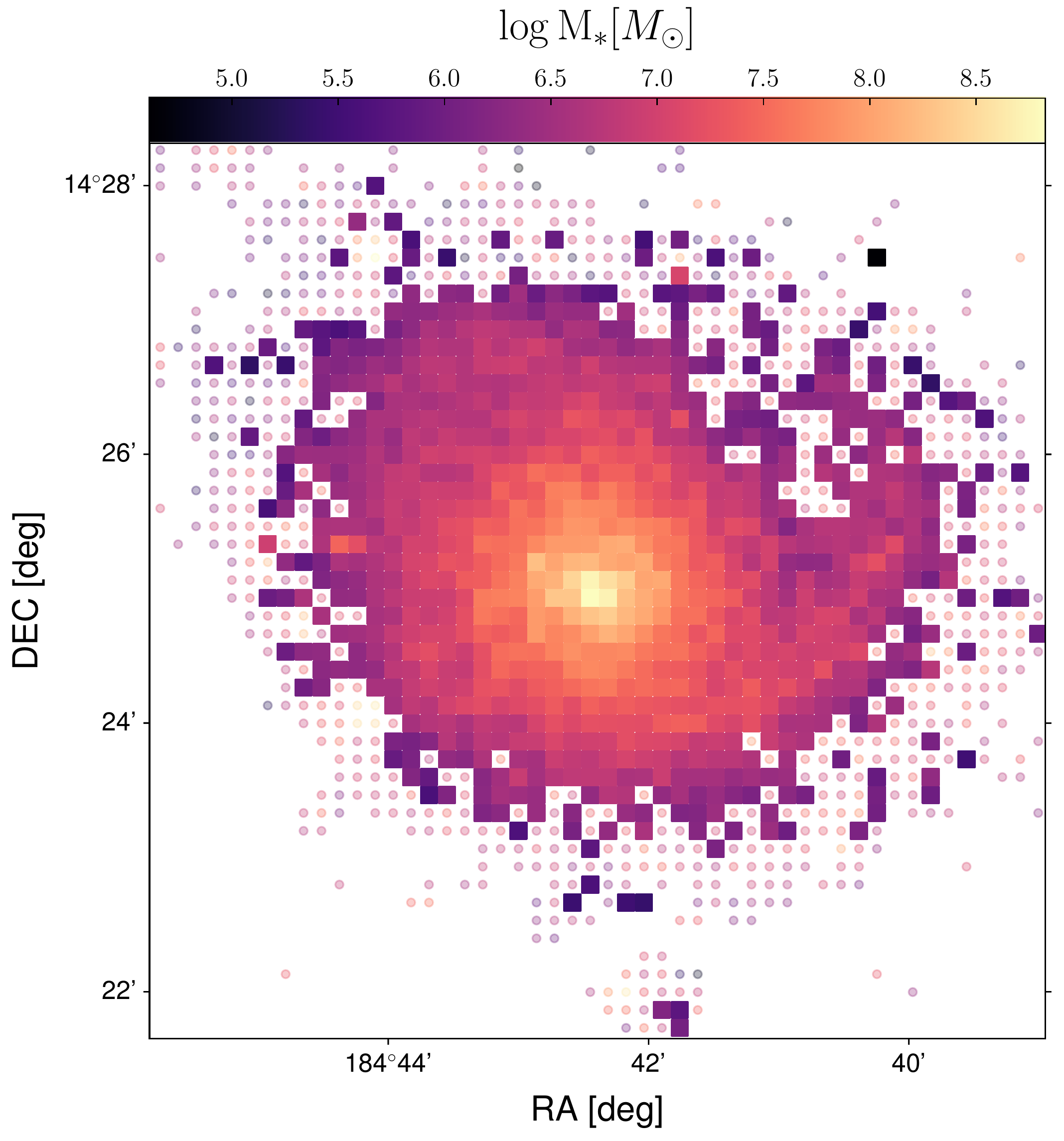}
	\includegraphics[width=1.02\columnwidth]{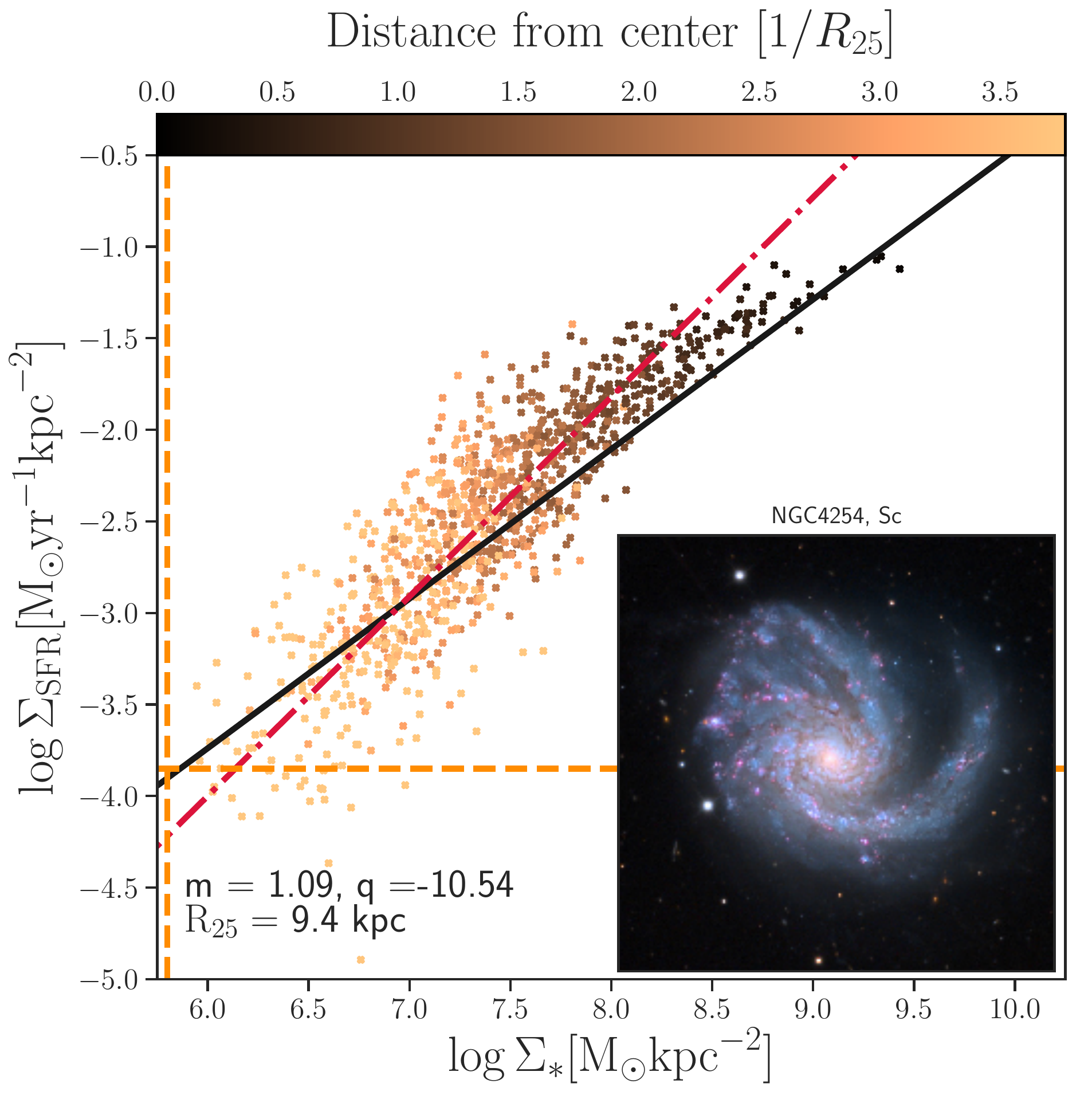}
    \caption{Same as Fig.\,\ref{fig:NGC0628}, for NGC4254.}
    \label{fig:NGC4254}
\end{figure*}

\begin{figure*}
	\includegraphics[width=\columnwidth]{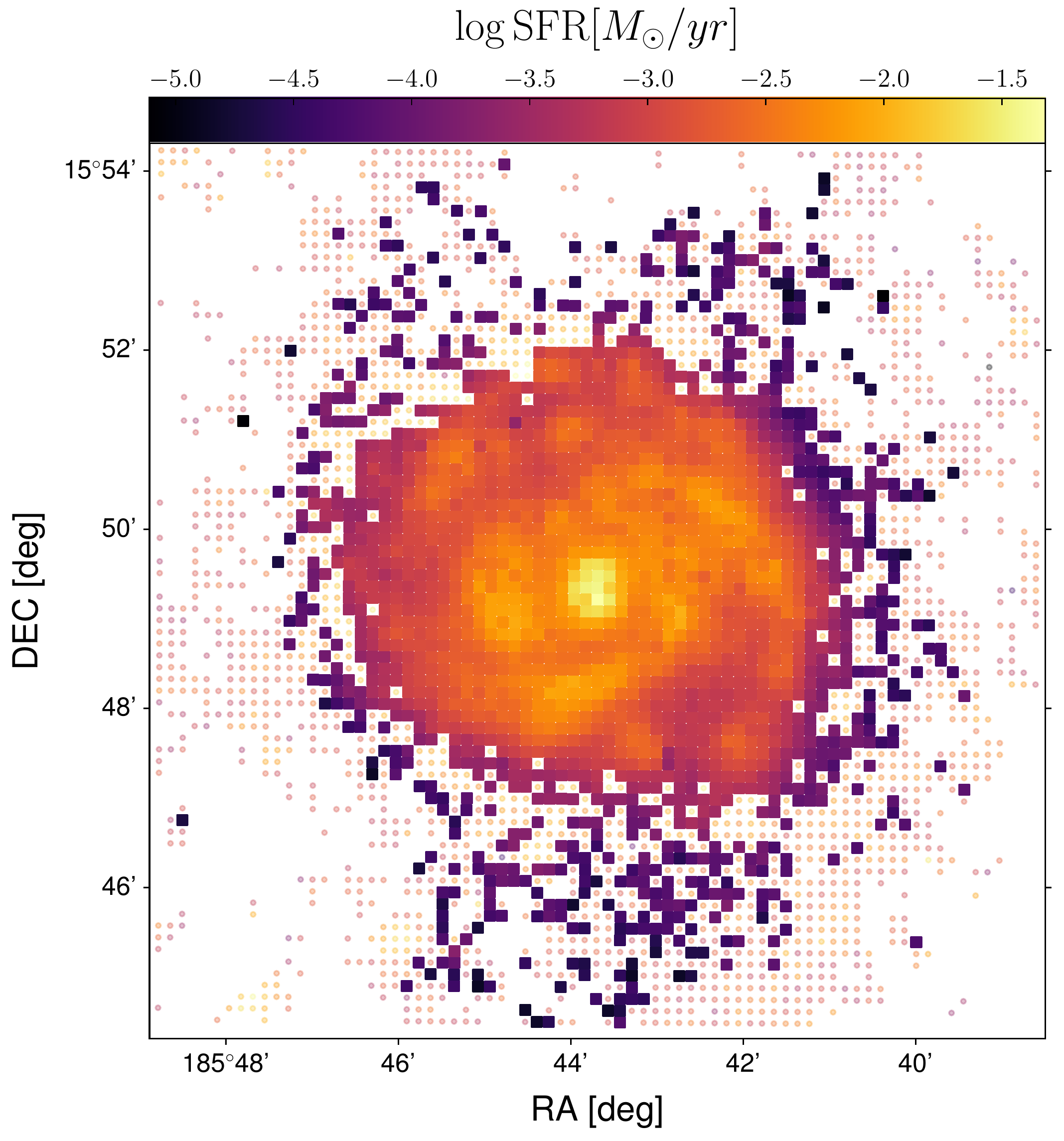}
	\includegraphics[width=\columnwidth]{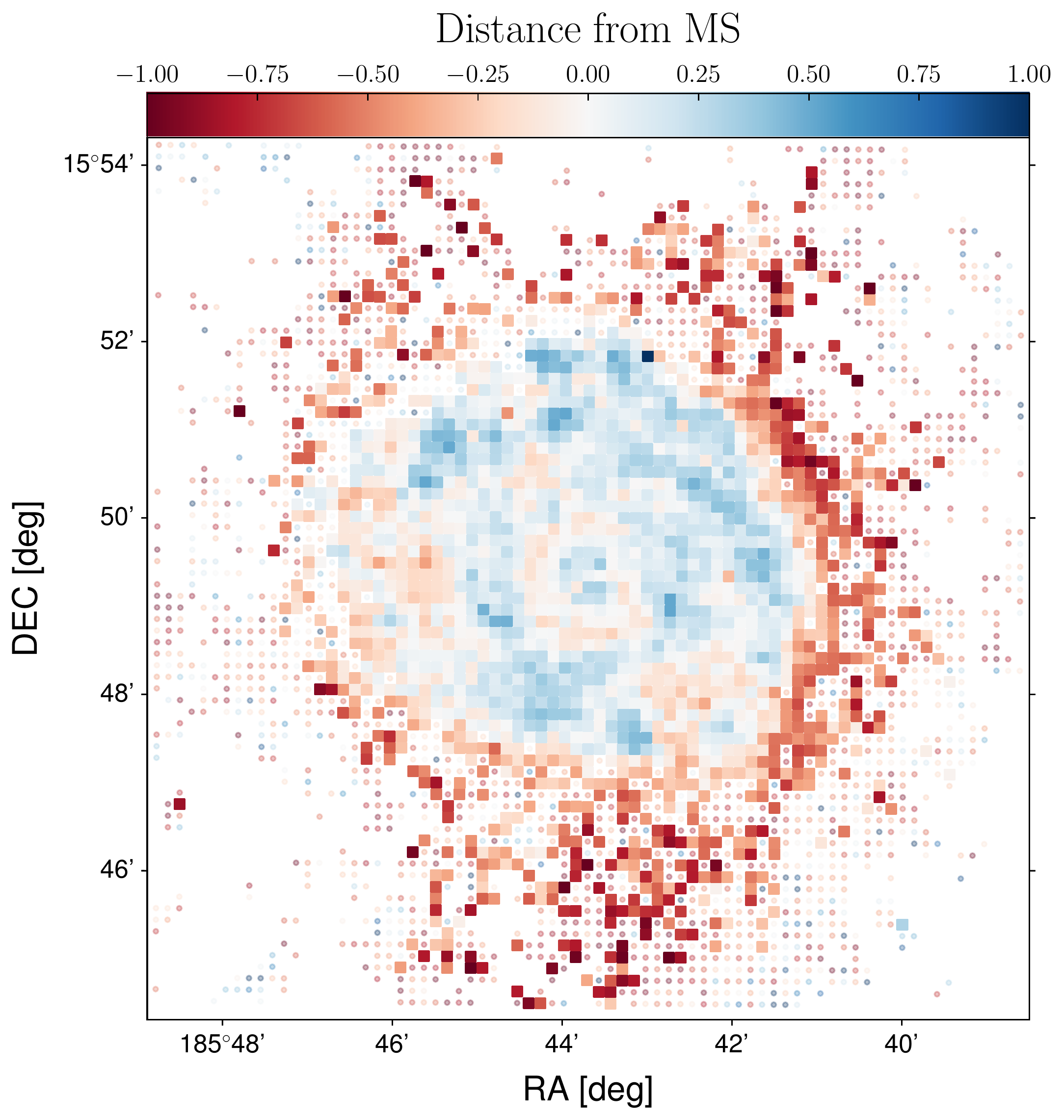}
	\includegraphics[width=\columnwidth]{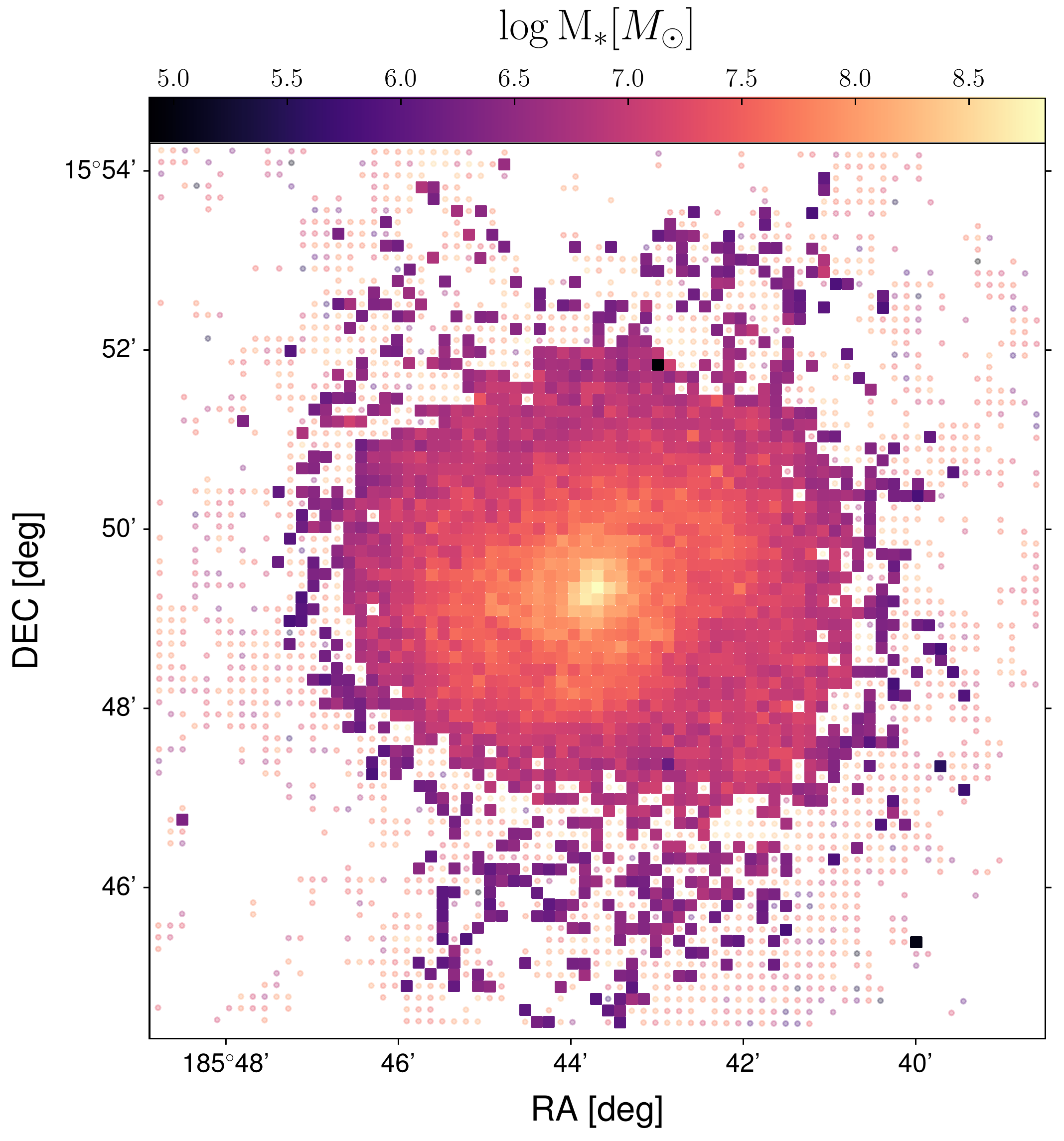}
	\includegraphics[width=1.02\columnwidth]{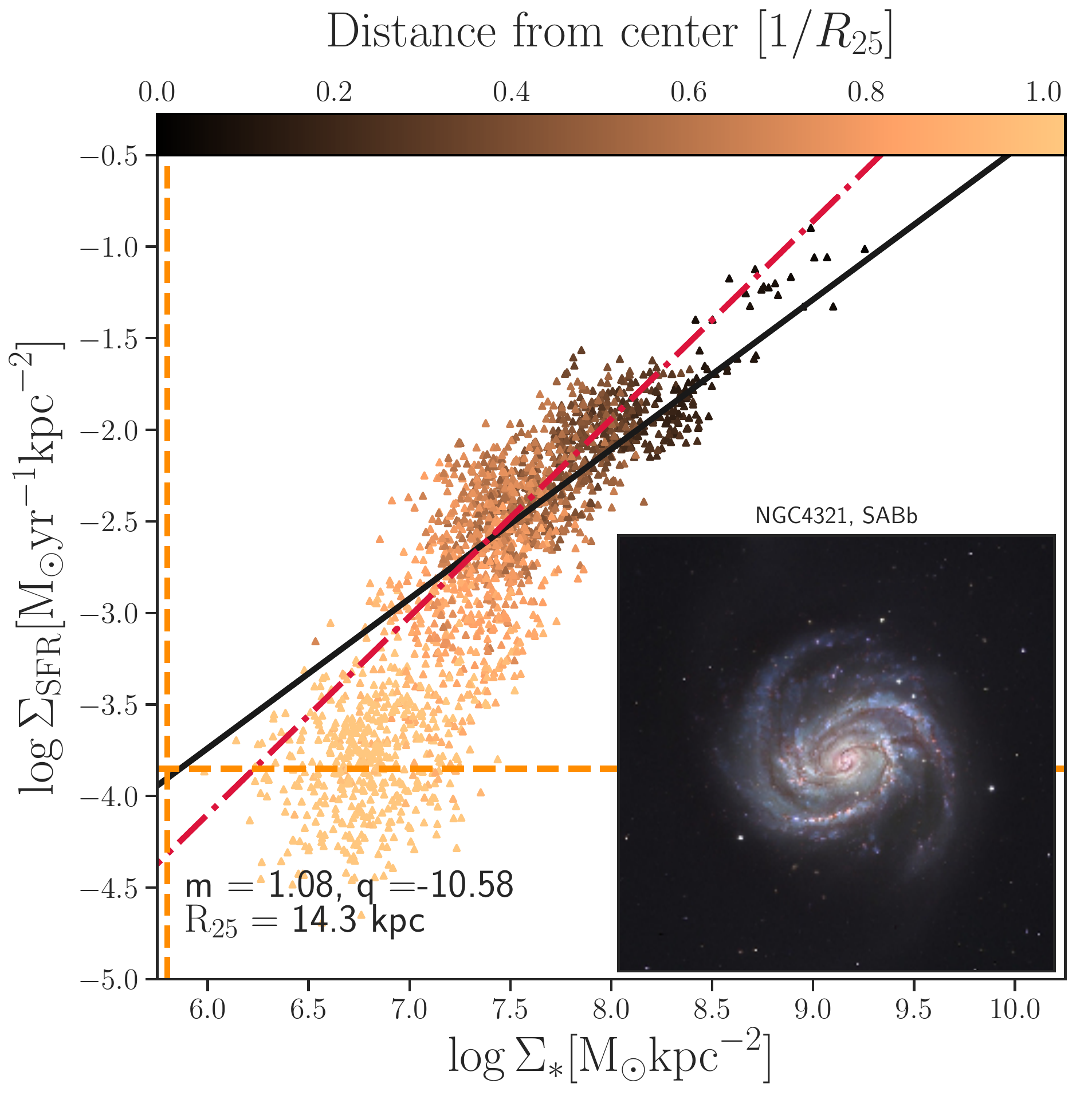}
    \caption{Same as Fig.\,\ref{fig:NGC0628}, for NGC4321.}
    \label{fig:NGC4321}
\end{figure*}

\begin{figure*}
	\includegraphics[width=\columnwidth]{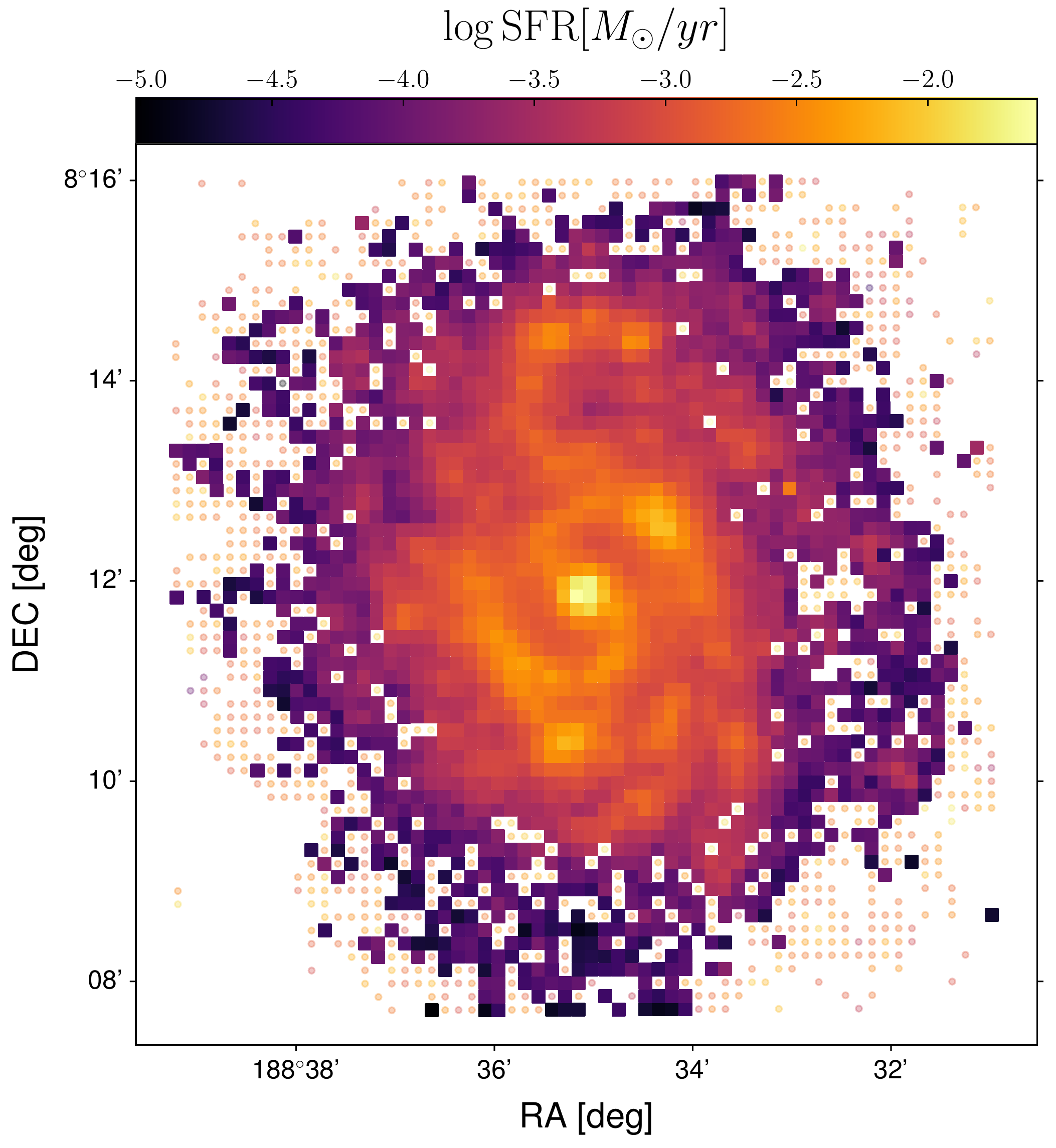}
	\includegraphics[width=\columnwidth]{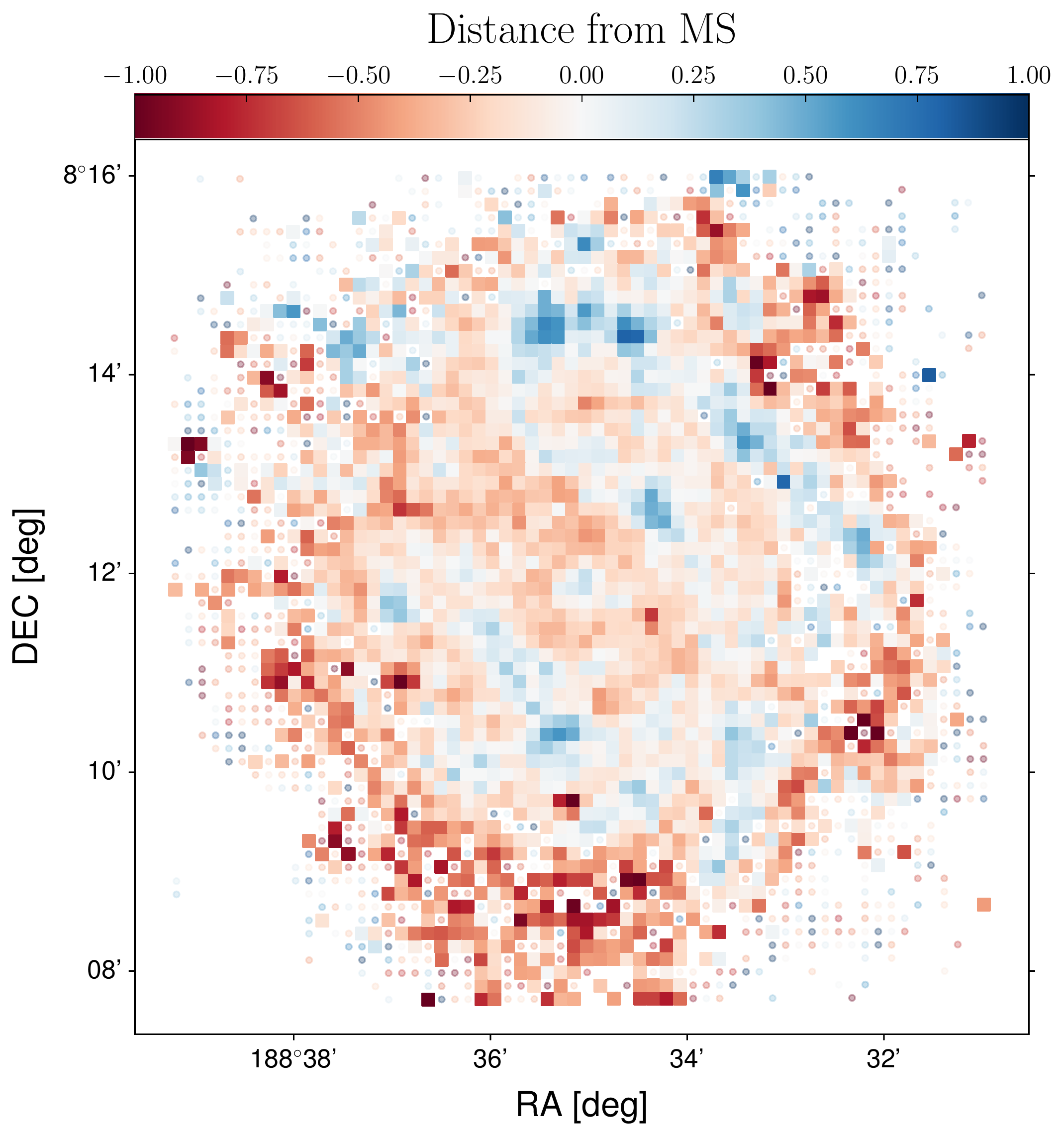}
	\includegraphics[width=\columnwidth]{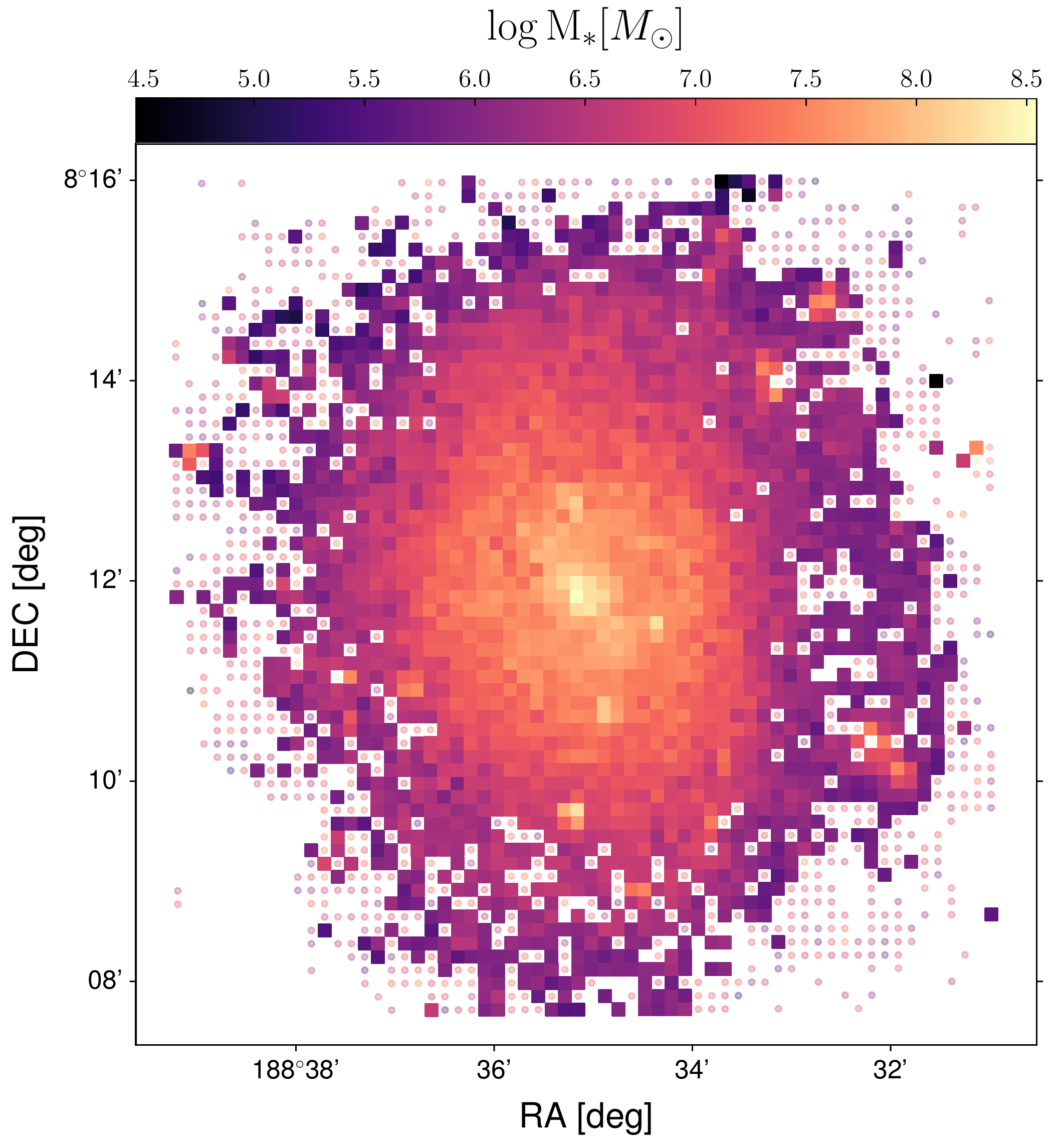}
	\includegraphics[width=1.02\columnwidth]{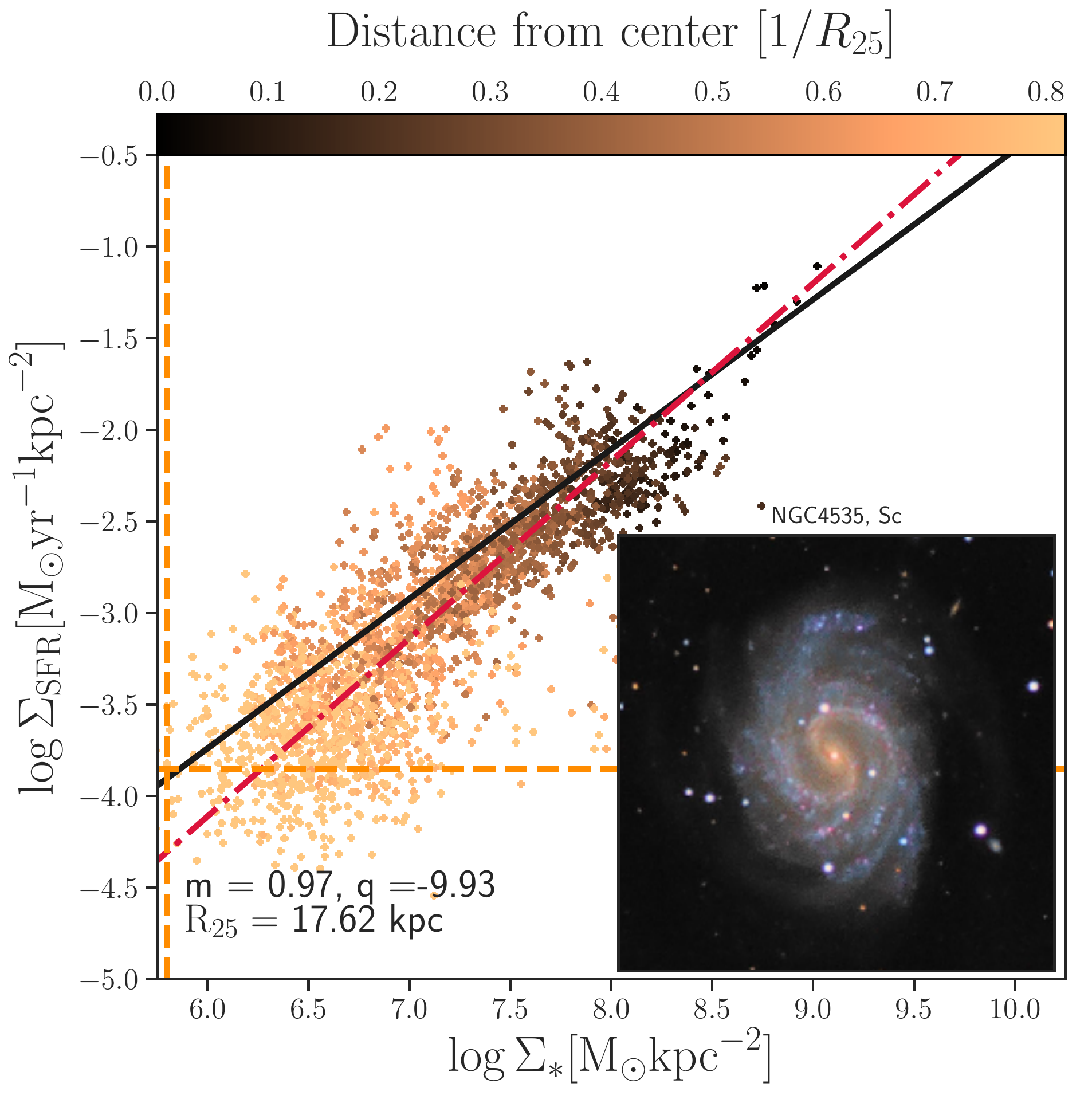}
    \caption{Same as Fig.\,\ref{fig:NGC0628}, for NGC4535.}
    \label{fig:NGC4535}
\end{figure*}

\begin{figure*}
	\includegraphics[width=\columnwidth]{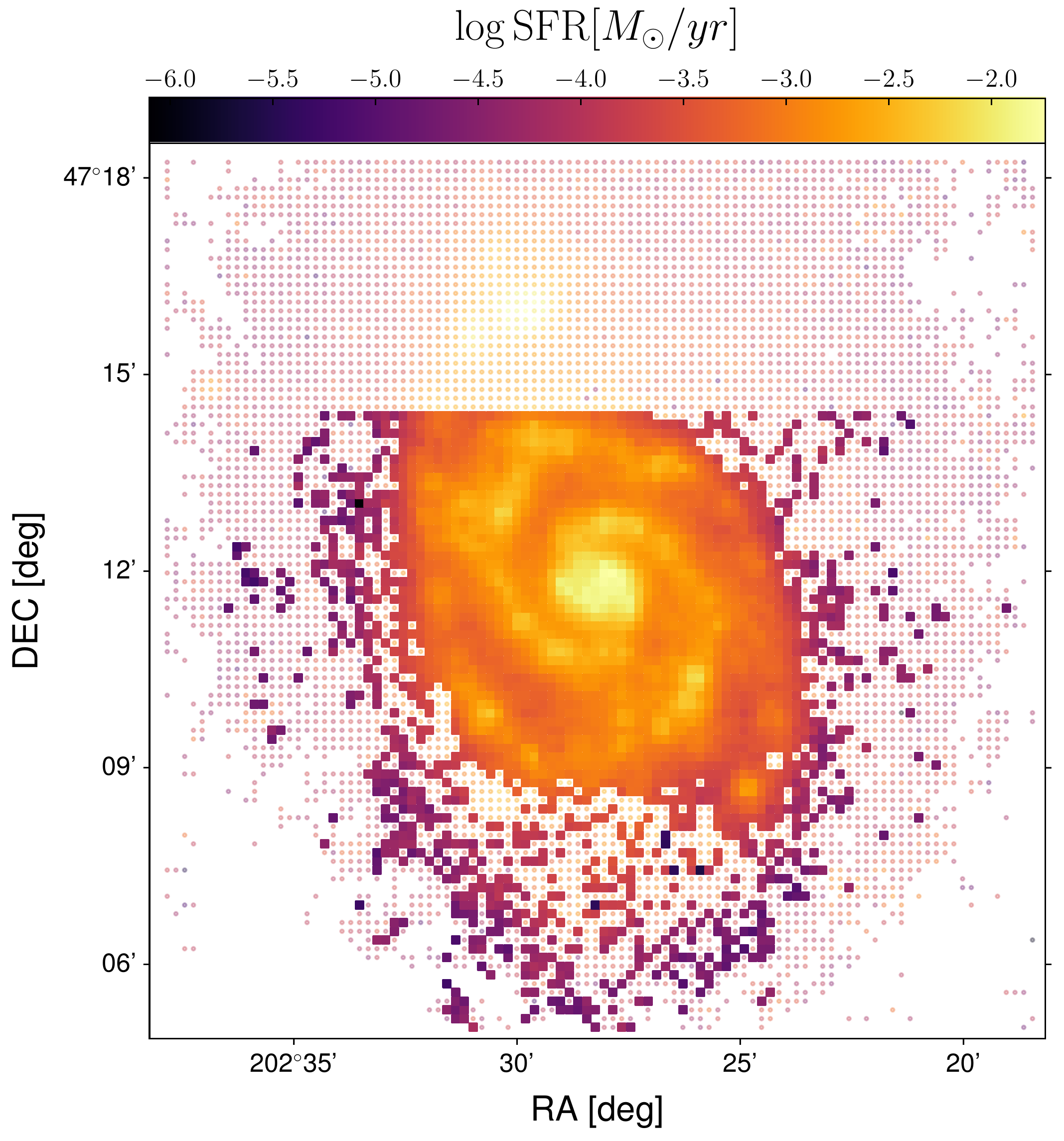}
	\includegraphics[width=\columnwidth]{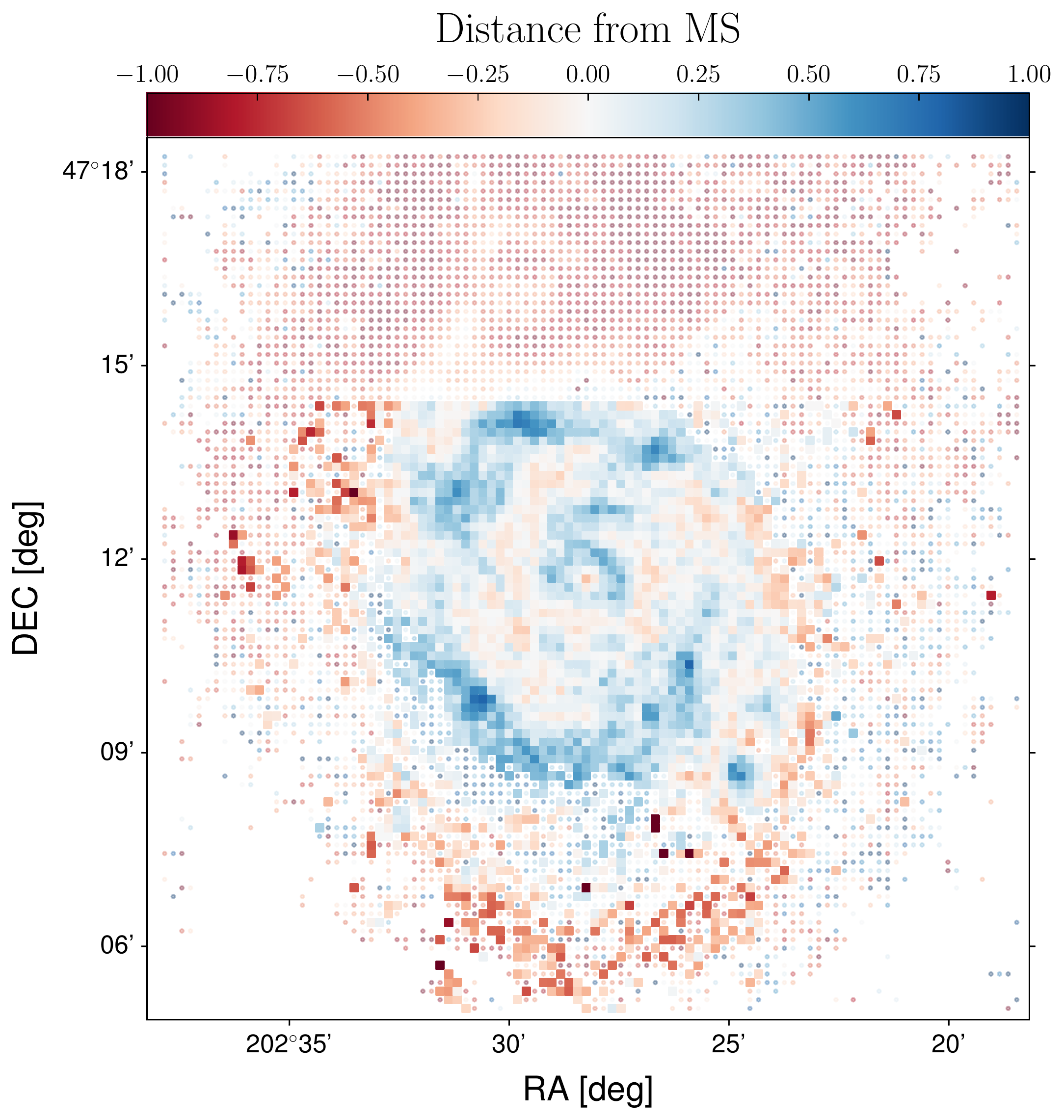}
	\includegraphics[width=\columnwidth]{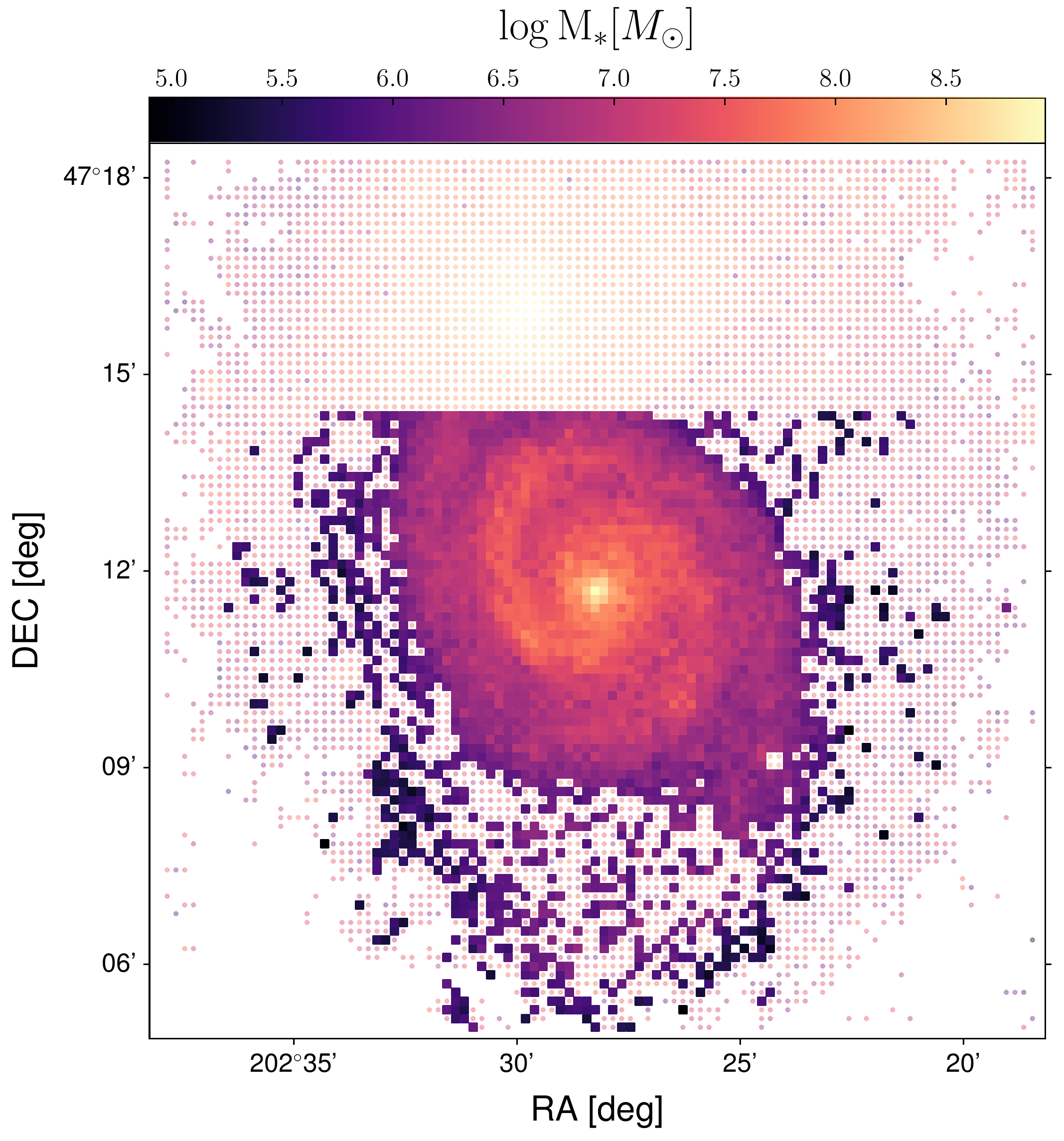}
	\includegraphics[width=1.02\columnwidth]{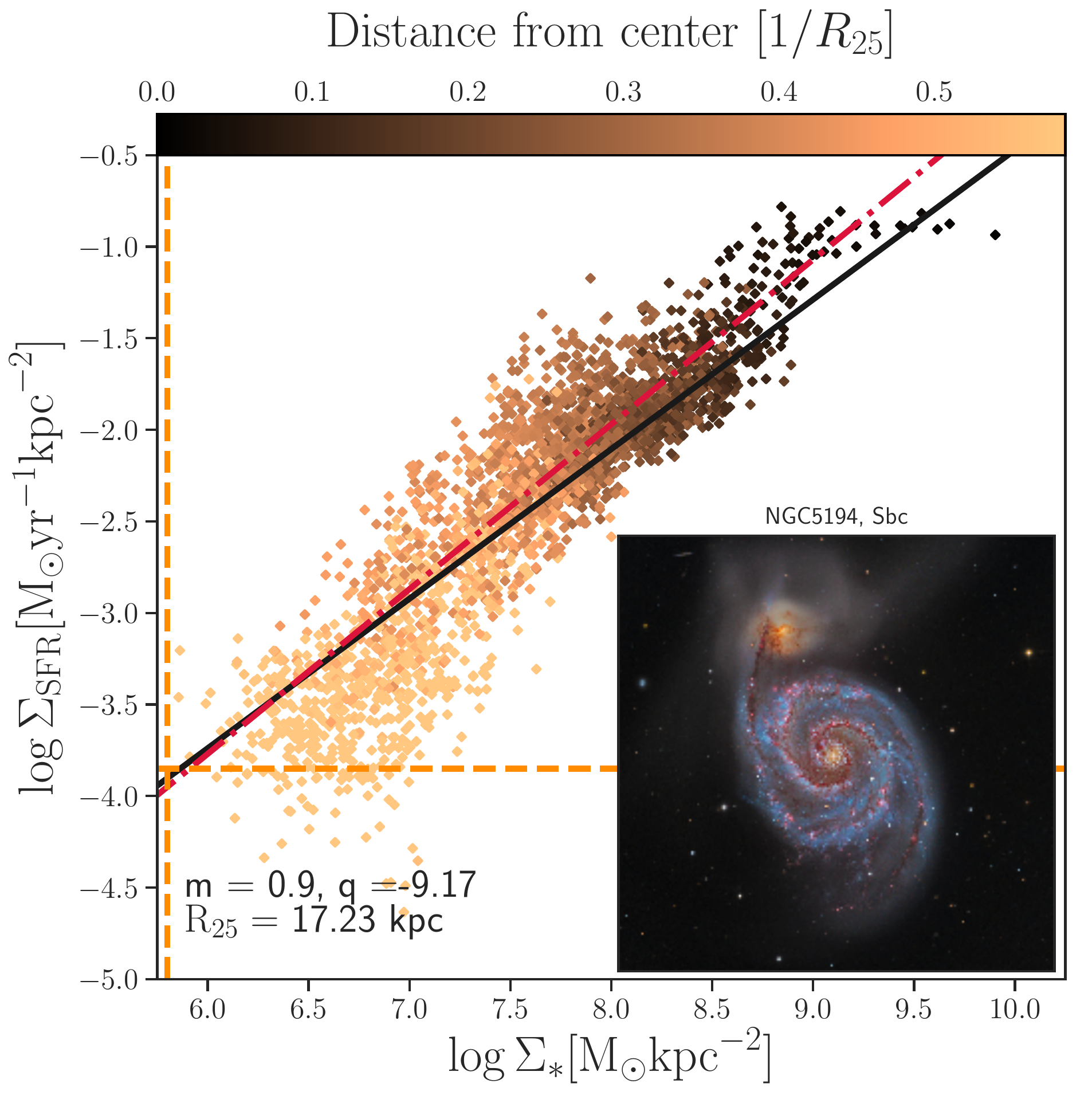}
    \caption{Same as Fig.\,\ref{fig:NGC0628}, for NGC5194.}
    \label{fig:NGC5194}
\end{figure*}

\begin{figure*}
	\includegraphics[width=\columnwidth]{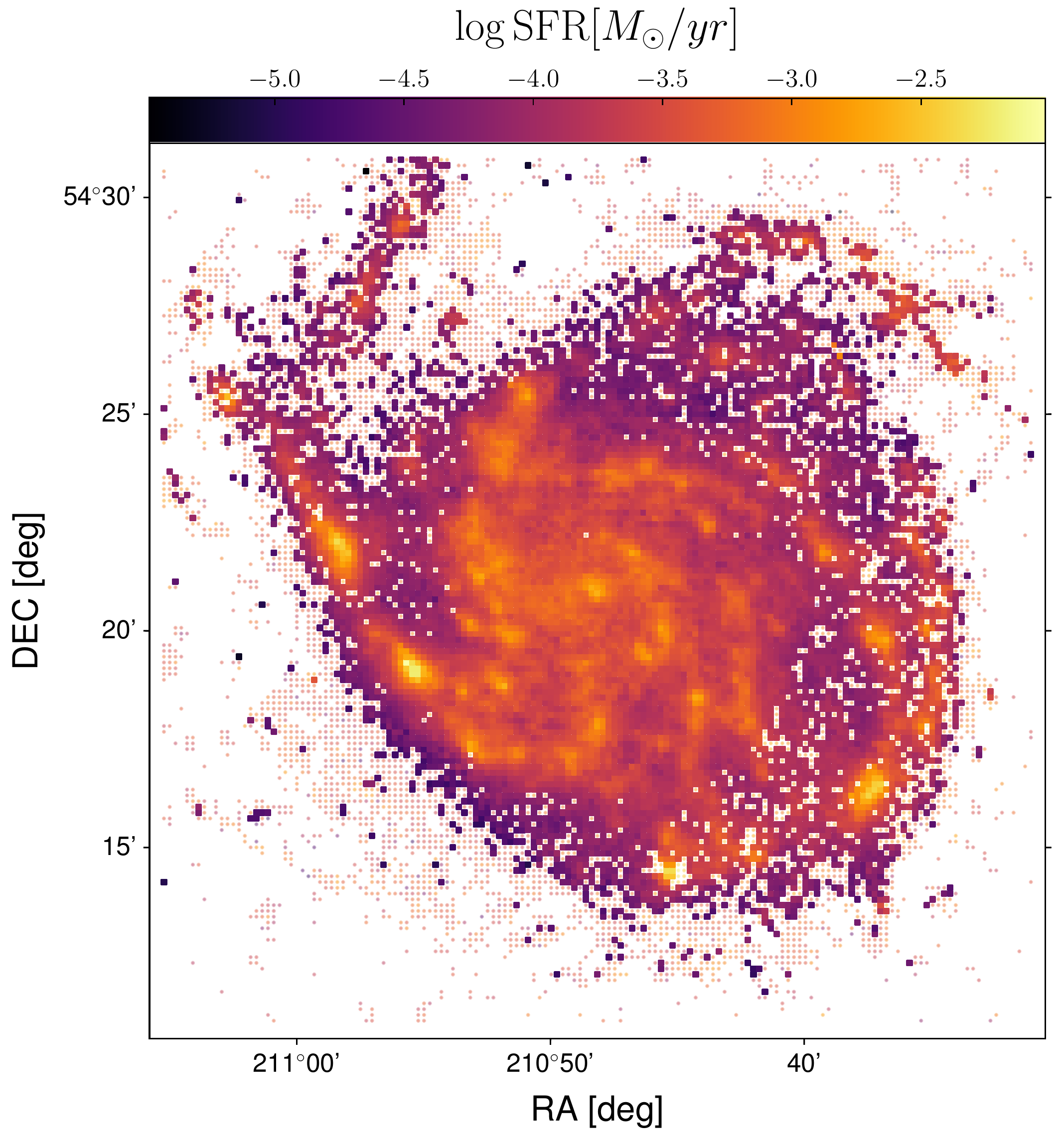}
	\includegraphics[width=\columnwidth]{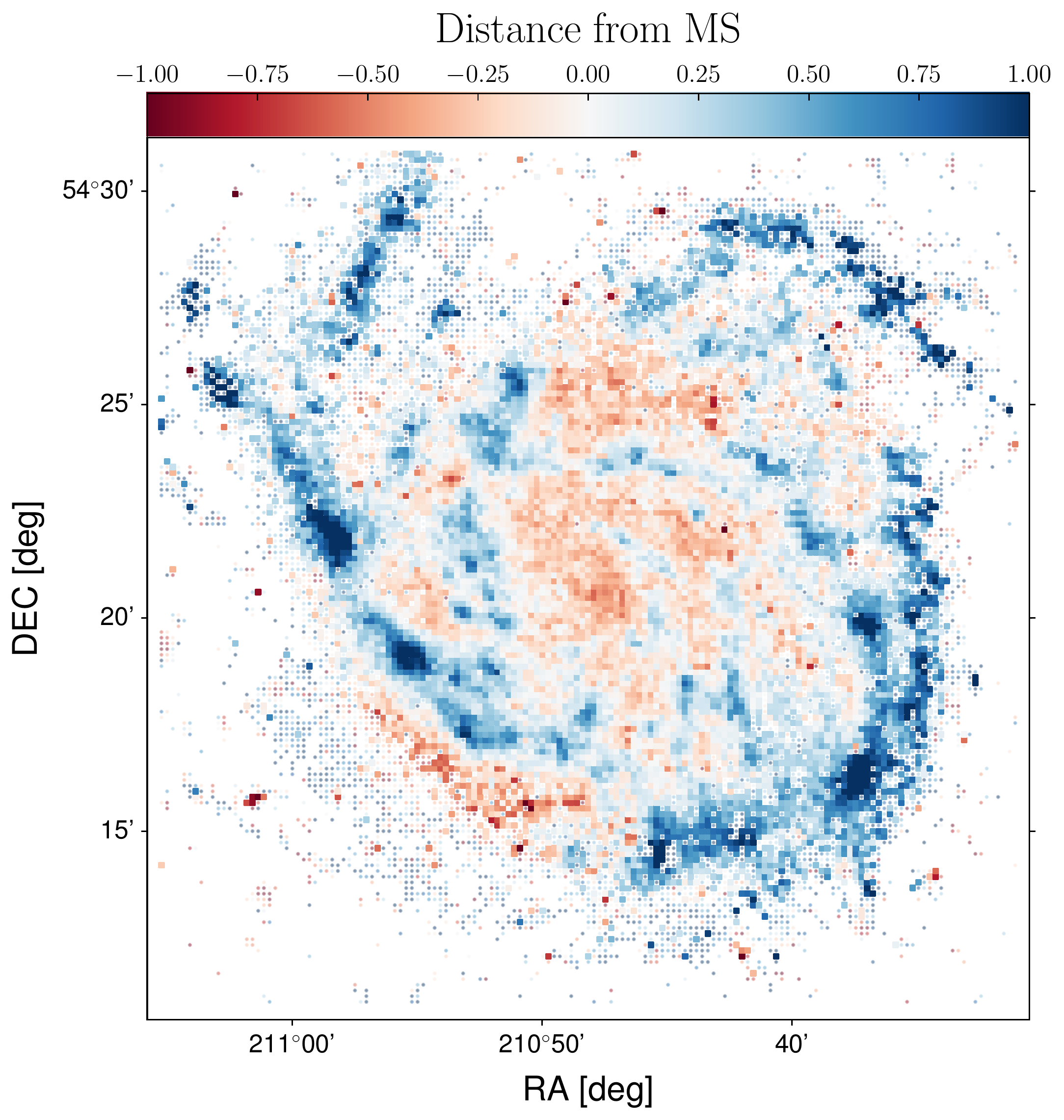}
	\includegraphics[width=\columnwidth]{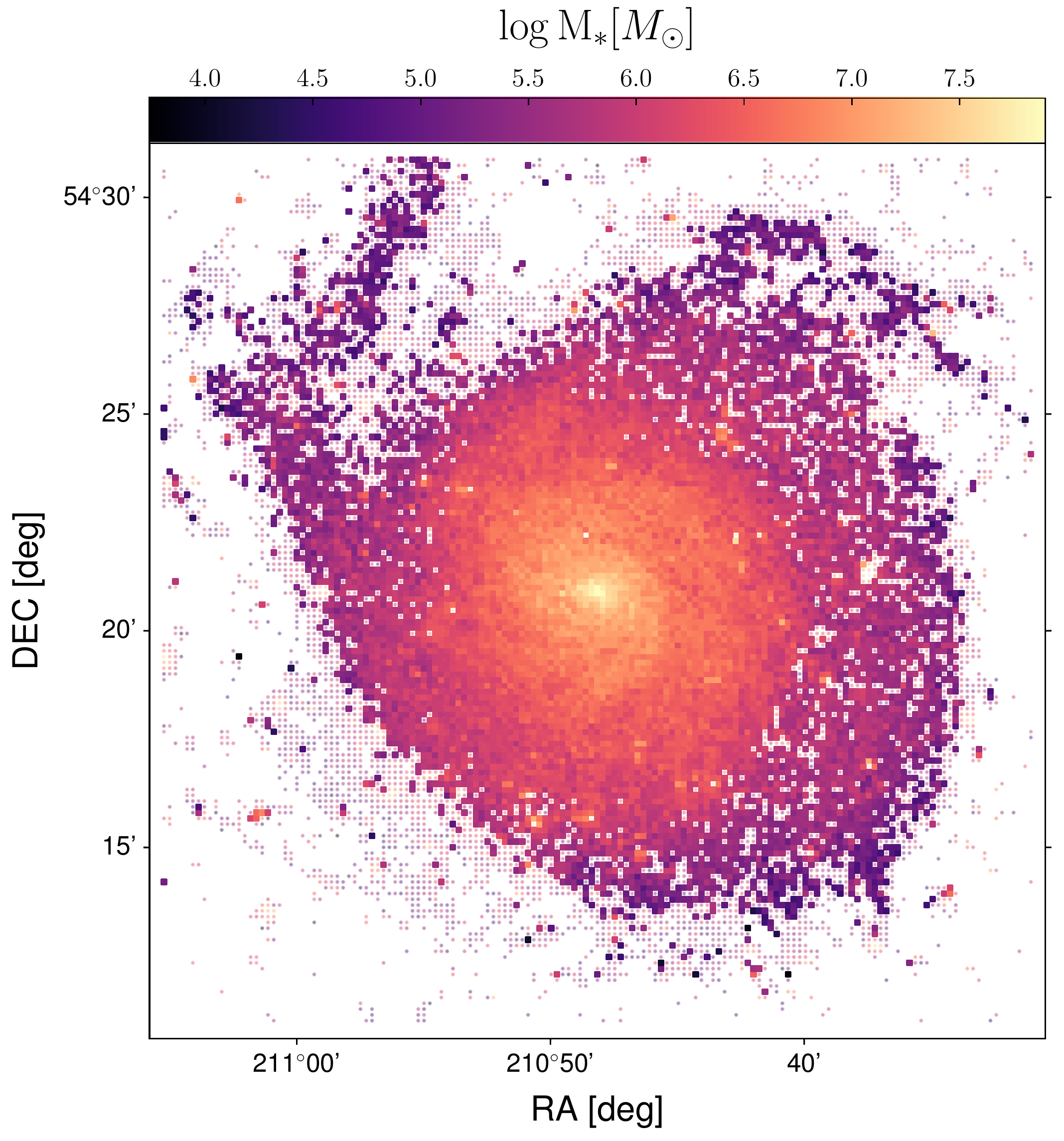}
	\includegraphics[width=1.02\columnwidth]{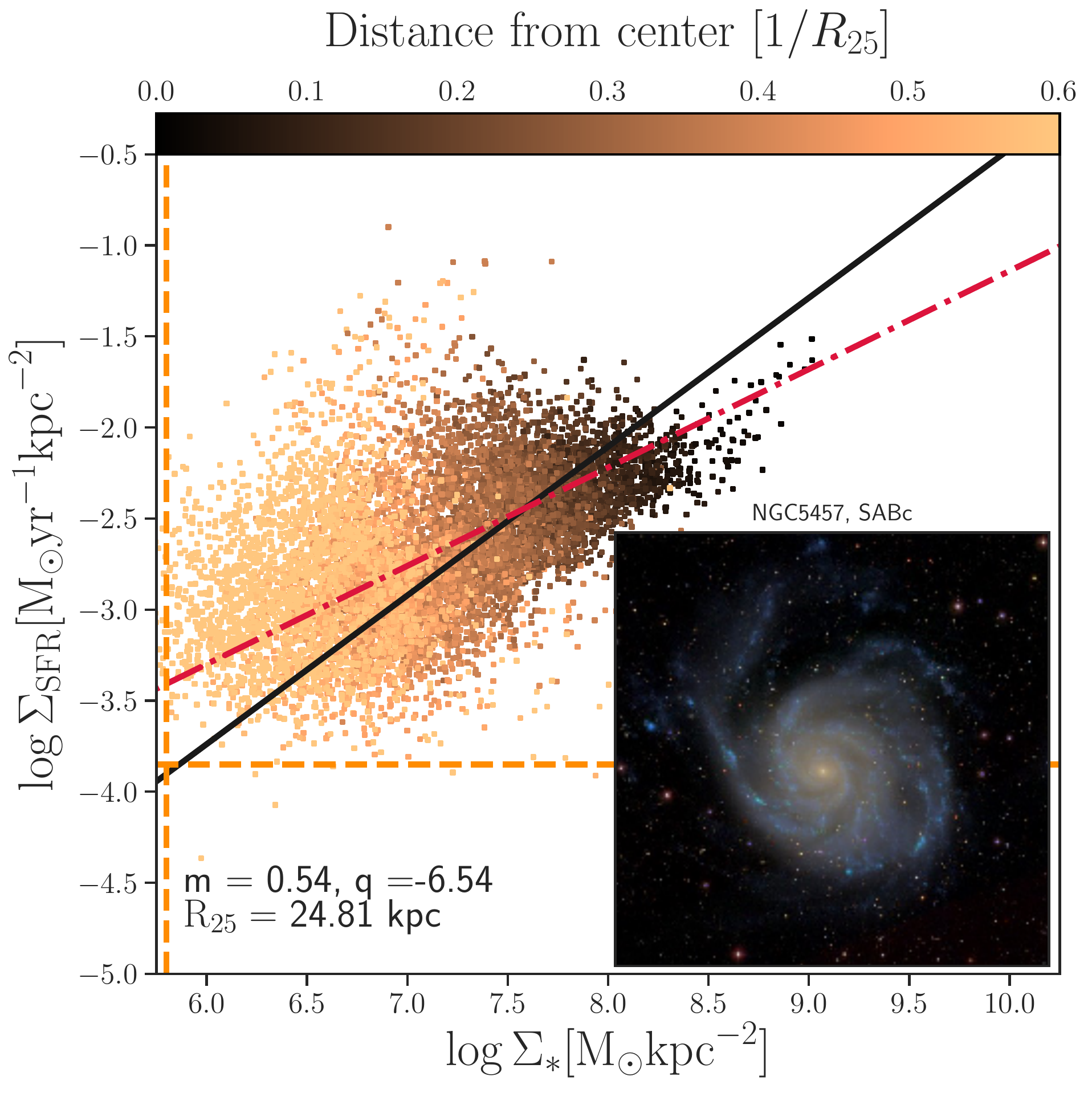}
    \caption{Same as Fig.\,\ref{fig:NGC0628}, for NGC5457.}
    \label{fig:NGC5457}
\end{figure*}



\bsp	
\label{lastpage}
\end{document}